\documentclass[aps, twocolumn,showpacs,floats,prd,superscriptaddress,nofootinbib]{revtex4-1} 
\usepackage{graphicx,amsmath,amssymb,amstext}
\usepackage{amssymb,amsbsy,amsfonts,amsthm,color}
\usepackage{epsfig}
\usepackage{graphicx}
\usepackage{subcaption}
\usepackage{sidecap}
\usepackage{floatrow}
\usepackage{multirow}
\usepackage{lastpage}
\usepackage{afterpage}
\usepackage{hyperref}
\usepackage[all]{hypcap}

\begin{document}

\title{\textbf{Screened fifth forces in parity-breaking correlation functions}}

\author{Darsh Kodwani}
\email{darsh.kodwani@physics.ox.ac.uk}
\affiliation{%
Department of Physics, University of Oxford, DWB, Keble Road, Oxford OX1 3RH, UK
}%

\author{Harry Desmond}
\email{harry.desmond@physics.ox.ac.uk}
\affiliation{%
Department of Physics, University of Oxford, DWB, Keble Road, Oxford OX1 3RH, UK
}

\date{\today}

\begin{abstract}
Cross-correlating two different types of galaxy gives rise to parity breaking in the correlation function that derives from differences in the galaxies' properties and environments. This is typically associated with a difference in galaxy bias, describing the relation between galaxy number density and dark matter density, although observational effects such as magnification bias also play a role. In this paper we show that the presence of a screened fifth force adds additional degrees of freedom to the correlation function, describing the effective coupling of the force to the two galaxy populations. These are also properties of the galaxies' environments, but with different dependence in general to galaxy bias. {We show that the parity-breaking correlation function can be calculated analytically, under simplifying approximations, as a function of fifth-force strength and the two populations' fifth-force charges, and explore the result numerically using Hu-Sawicki $f(R)$ as a toy model of chameleon screening.} We find that screening gives rise to an octopole, which, in the absence of magnification bias, is not present in any gravity theory without screening and is thus a qualitatively distinct signature. The modification to the dipole and octopole can be $\mathcal{O}(10\%)$ and $\mathcal{O}(100\%)$ respectively at redshift $z \gtrsim 0.5$ due to screening, but decreases towards lower redshift. The change in the background power spectrum in $f(R)$ theories induces a change in the dipole of roughly the same size, but dominant to the effect of screening at low $z$. While current data is insufficient to measure the parity-breaking dipole or octopole to the precision required to test these models, future surveys such as DESI, Euclid and SKA have the potential to probe screened fifth forces through the parity breaking correlation function. 
\end{abstract}

\maketitle

\section{Introduction}
\label{sec:intro}

The standard cosmological model, $\Lambda$ Cold Dark Matter ($\Lambda$CDM), relies on the assumption that gravity is described by a rank 2 symmetric tensor on large scales. The cosmological parameters fitted to observations such as the cosmic microwave background \cite{Adam:2015rua, Hinshaw:2012aka} and large scale structure \cite{Ivezic:2008fe, Amendola:2016saw, DESI} assume General Relativity (GR) as the theory of gravity. However, fundamental puzzles in the $\Lambda$CDM model such as the nature of dark energy and dark matter have encouraged research into the cosmological implications of modifications to GR, dubbed modified gravity. There are a plethora of modified gravity models, ranging from the addition or alteration of terms in the Einstein-Hilbert action to the explicit coupling of additional scalar, vector or tensor fields (see e.g. \cite{Clifton:2011jh} for a review). This motivates expanding the parameter space of traditional cosmological inferences, as well as designing novel probes with maximum sensitivity to new gravitational degrees of freedom.

The diversity of modified gravity models makes it inconvenient to test them individually. This has led to the development of generalised frameworks within which many theories may be tested simultaneously, for example the Effective Field Theory of Dark Energy \cite{Gubitosi:2012hu} and Parametrised Post-Friedmannian framework \cite{Ferreira:2014mja}.
Modified gravity theories may also be characterised by the \emph{screening mechanisms} they incorporate to hide the effects of new interactions at small scales. As almost all viable theories employ one of just a handful of screening mechanisms, probing such a mechanism is tantamount to probing a potentially broad class of theories. These theories may often be cast as screened scalar--tensor theories (see \cite{Khoury_rev, Jain_rev, Joyce_rev} and references therein), in which a long-range dynamical field couples universally to matter, generating a new (``fifth'') force between masses. The Lagrangian is designed so that the strength or range of the force depends on the local gravitational environment: the fifth force is suppressed in high-density regions (such as within the Solar System where the most stringent constraints exist; see \cite{Adelberger:2003zx} and references therein) but emerges at the lower densities of the Universe at large. This leads to differences in both inter-galaxy clustering and intra-galaxy morphology and dynamics between galaxies in stronger vs weaker gravitational fields. The latter class of signal has been the subject of a range of tests in recent years \cite{Hui, Jain_Vanderplas, Cepheid, Vikram, Vikram_RC, Desmond_1, Desmond_2, Desmond_3}; our purpose here is to explore the former.

This paper investigates the galaxy correlation function (CF) as a probe of screened fifth forces. It is well known that standard general relativistic effects in large scale structure give rise to odd CF multiples \cite{Bonvin:2013ogt, Croft:2013taa} (or, in Fourier space, an imaginary part of the power spectrum \cite{McDonald:2009ud, Yoo:2012se}). These CFs break two distinct symmetries, at different scales and with different physical causes:

\begin{enumerate}

\item In a cluster environment, the CF is asymmetric under a swapping of spatial locations of galaxies along the line of sight. This is because galaxies behind the centre of a deep potential well appear closer in redshift space (and therefore more strongly correlated) than galaxies in front, due to the redshift induced by the gravitational potential.\footnote{See fig. 2 of \cite{Bonvin:2013ogt}.} This constitutes a breaking of the spatial isotropy symmetry group SO(3) into an SO(2) perpendicular to the line of sight, and is present for a \emph{single} population of galaxies on cluster scales $\sim1-10$ Mpc \cite{PhysRevLett.114.071103, cluster_test}.

\item The second type of symmetry breaking is present only in the cross-correlation of \emph{two different} populations of galaxies: $\langle \Delta_B(\vec{x}_1) \Delta_F(\vec{x}_2) \rangle \neq \langle \Delta_F(\vec{x}_1) \Delta_B(\vec{x}_2) \rangle $. Here $B$ and $F$ denote ``bright" and ``faint" galaxies, meaning that they trace the underlying matter field in different ways or otherwise have different properties pertinent to their clustering. For example, even if the galaxies form within the same dark matter density field and hence gravitational potential, they may form at different rates and hence end up with different final number densities. This is manifest in a difference in their bias. These differences in their spatial statistics correlate most strongly with their $z=0$ halo masses \cite{Kaiser, Sheth, Tinker}, with secondary effects deriving from other galaxy and halo properties (``assembly bias''; e.g. \cite{Gao, Wechsler, Mao}). Thus galaxy subsamples that differ in any observable that correlates with halo mass, e.g. luminosity or type, will manifest a parity-breaking CF. The formation of these different types of galaxy may be driven by the tidal field or other features of the cosmic web, making the effect a function of large-scale environment. This effect is present on larger scales than isotropy violation, $\sim$100 Mpc, and breaks the qualitatively different symmetry group $\mathbb{Z}_2$, which is parity under swapping the discrete $B$ and $F$ labels.

Although bias is the galaxy property conventionally responsible for giving the two populations different clustering, another possibility is sensitivity to a screened fifth force. Screened theories of gravity produce modifications to the gravitational force that depend on galaxies' internal properties and environments (Sec.~\ref{sec:screening}), in such a way that lower mass galaxies in lower density regions effectively feel stronger gravity. As we describe in detail in Sec.~\ref{sec:corr}, this alters the Euler equation and hence the number densities predicted by relativistic perturbation theory. We illustrate the effect schematically in Fig.~\ref{windy} where we show two galaxies at different spatial locations $x_1$ and $x_2$ in the same gravitational potential, but with different biases and screening parameters due to their different \emph{environments}. The environment is represented by the tree, which experiences wind at $x_2$ but not at $x_1$, delineating the fact that the environments are different at the two locations.

\begin{figure}
\begin{center}
\includegraphics[scale = 0.25]{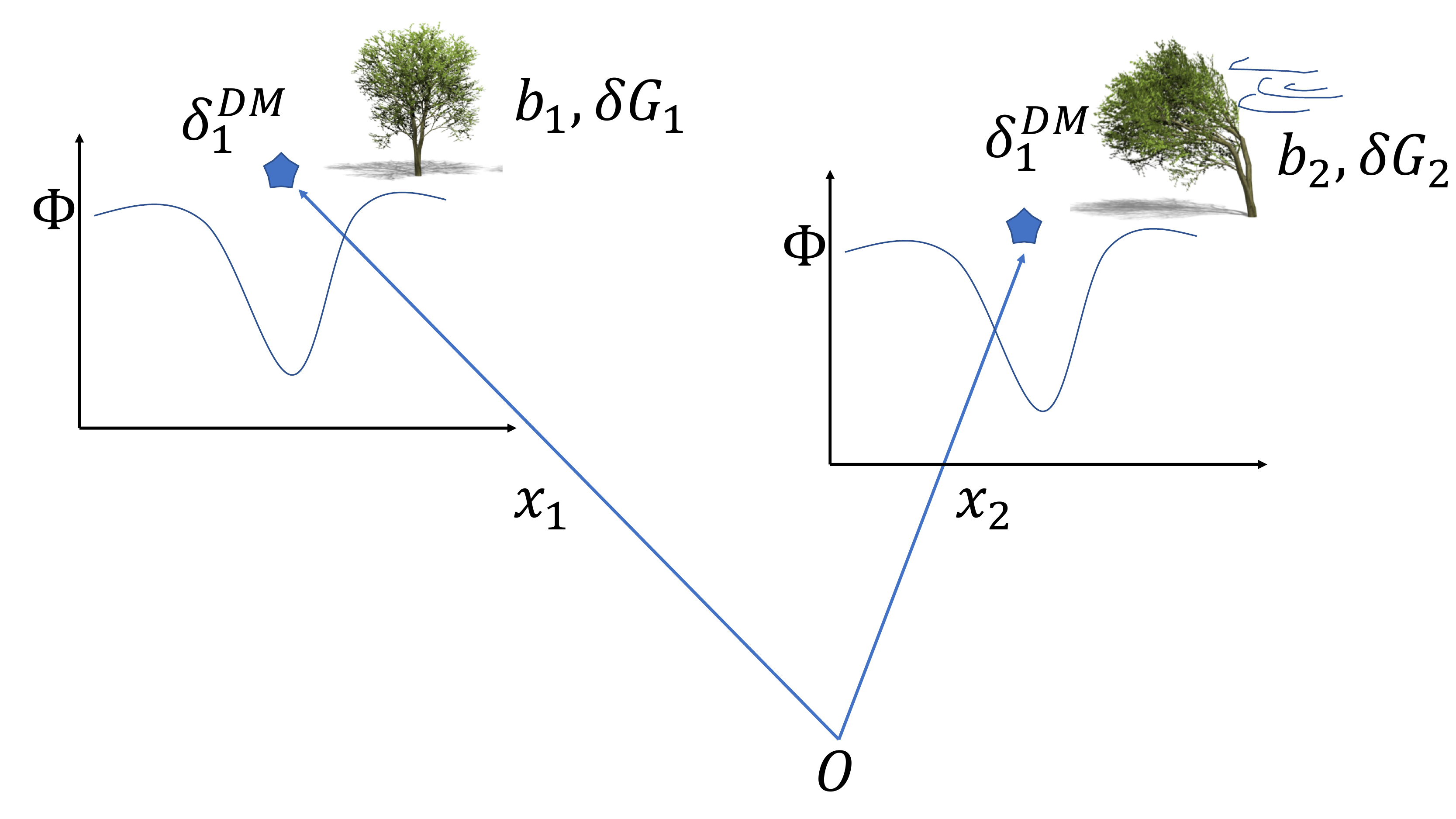}
\caption{
Schematic of the effect of parity breaking in the CF. The two galaxies are located in regions of identical external dark matter density and hence gravitational potential (i.e. excluding the potential due to the galaxies' halos themselves, which are responsible for their self-screening), but the larger-scale environments are different. This is illustrated by the surroundings of the trees, which are windy at $x_2$ but not $x_1$. This difference in environment affects various properties of the galaxy that impact its clustering, including its bias, magnification bias and, of particular interest here, sensitivity to a fifth force (quantified by $\delta G \equiv \Delta G/G_\text{N}$). Swapping these parameters between the galaxy overdensities at $x_1$ and $x_2$ alters the CF, so that it is not symmetric in the galaxy labels.}
\label{windy}
\end{center}
\end{figure}

\end{enumerate}

Both of these effects are proportional to the gradient of the gravitational potential and hence receive contributions from fifth forces. In particular, as noted in \cite{Bonvin:2013ogt, Bonvin:2018ckp}, in GR the gravitational redshift term of the parity-breaking CF is precisely cancelled by the light-cone term and part of the Doppler term. This apparent coincidence is a result of the equivalence principle, whereby both light and matter feel the same potential. In contrast, theories that are conformally equivalent to GR yet include fifth forces effectively violate the equivalence principle due to the effect of the fifth force on timelike but not null geodesics. This reintroduces the redshift term, enabling relativistic effects to provide a consistency check on the validity of the Euler equation. While this effect would also be present under an unscreened fifth force, the additional effect of screening is that the two types of galaxy that enter the CF may feel \emph{different} fifth-force strengths due to their different gravitational environments endowing them with different scalar charges.\footnote{The screened fifth force that we invoke does not have to be mediated by a scalar field, but we will assume so for simplicity.} Thus timelike geodesics are affected differentially by the fifth force as a function of their trajectory, in further violation of the strong as well as weak equivalence principle. We show in Sec.~\ref{sec:corr} that this introduces interesting novel behaviour into the parity-breaking CF. We emphasise that this is not a fundamental breaking of either parity or the equivalence principle at the level of the Lagrangian. Rather, it is an effective violation stemming from differences in galaxy properties.

The structure of this paper is as follows. In Sec.~\ref{sec:screening} we provide theoretical background on screened fifth forces, and in Sec.~\ref{sec:corr} we lay out the formalism for calculating the CF in their presence, paying particular attention to the relativistic parity-breaking part. Sec.~\ref{sec:results} presents our results for the dipole and octopole, specialising to a specific chameleon-screened theory, Hu-Sawicki $f(R)$, when numerical results are required. Sec.~\ref{sec:conc} provides a summary of our results and a brief discussion of future work, including prospects for testing the effects observationally. Appendix~\ref{corr_funcs} presents the full calculation of the CF, and Appendix~\ref{horndeski} shows how our work could be made more general within the chameleon paradigm by casting the chameleon action in Horndeski form.

\section{Screened fifth forces}
\label{sec:screening}

To see the need for screening in theories with new dynamical degrees of freedom, consider the behaviour of a free light scalar $\phi$. 
{The Klein-Gordon equation for the scalar field in the quasi-static limit is}
\begin{equation}
	\nabla^2 \phi = 8 \pi \rho G \alpha, \label{f_Poisson}
\end{equation}
where $\rho$ is the energy density and $\alpha$ the coupling coefficient of the scalar field to matter. This is solved by $\phi = 2\alpha G M/r$, which produces the fifth force
\begin{equation}
        F_5 = -\alpha \nabla \phi = -2 \alpha^2 G M / r^2 = 2 \alpha^2 F_N. \label{eq:F5}
\end{equation}
This modifies the spatial part of the weak-field metric but not the temporal part, making it sensitive to tests of the Parametrised Post-Newtonian light-bending parameter $\gamma$. The most stringent constraint derives from the radio link to the Cassini spacecraft, which requires $\gamma < \mathcal{O}(10^{-5})$ and hence $\alpha \le \mathcal{O}(10^{-3})$ \cite{Bertotti:2003rm}. In a Friedman-Robertson-Walker (FRW) metric, the cosmological effect of the scalar field is given by
\begin{equation}
	{\phi}'' + 3 H{\phi}' + \alpha G \rho = 0,
\end{equation}
{where $H$ is the Hubble parameter and prime denotes derivative with respect to cosmic time}. For values of $\alpha$ this small, the final term is negligible and hence the fifth force is a tiny perturbation to GR dynamics.

The operation of screening may be seen from the most general equation of motion the scalar field could obey:
\begin{equation}
	\Sigma^{ij}(\phi_0) \partial_i \phi \partial_j \phi + m^2_\text{eff}(\phi_0) \phi = 8 \pi G \alpha(\phi_0) \rho \label{gen_poisson},
\end{equation}
where $\Sigma^{ij}$ is a matrix that allows for general non-linear and non-diagonal kinetic terms. The effective mass is given by the second derivative of the potential term in the Lagrangian, and $\alpha$ is in general a function of the background field value $\phi_0$. The solution to this equation is
\begin{equation}
	\phi = 2 \alpha(\phi_0) G \frac{M}{|\Sigma(\phi_0)|r} e^{-m_\text{eff}(\phi_0)r},
\end{equation}
which illustrates the three qualitatively different mechanisms for removing the influence of the scalar field in high-density regions:
\begin{enumerate}
\item The field can be made short range by giving it a large effective mass at high density, i.e $m_\text{eff} \gg 1/R$ where $R$ is the size of the system. This is called \emph{chameleon screening} \cite{Khoury:2003rn}.
\item The amplitude of the force can be decreased by reducing the coupling to matter $\alpha(\phi_0)$. This is most commonly done via spontaneous breaking of a $\mathbb{Z}_2$ symmetry in the field configuration, in which case it is known as \emph{symmetron screening} \cite{Hinterbichler:2010es, Clampitt:2011mx}. 
\item The amplitude of the nonlinear kinetic terms can be increased, $\Sigma^{ij} \gg 1$, effectively decoupling the scalar field from matter. Depending on the precise implementation this is called \emph{kinetic} \cite{kinetic} or \emph{Vainshtein} screening \cite{Vainshtein}.
\end{enumerate}
Under any one of these mechanisms, the scalar charge of an object depends on its density and, in the case of chameleon or symmetron screening, also on its gravitational environment \cite{Brax, Falck, Zhao}. Low mass unscreened objects feel the full fifth force, which, in case the scalar field is light relative to their size, effectively causes them to feel an enhanced Newton's constant $G = G_\text{N} + \Delta G$. Conversely, high mass screened objects have no scalar charge and hence decouple from the fifth force and feel regular gravity. We describe this with a parameter $\delta G \equiv \Delta G/G_\text{N}$, which takes the value $2 \alpha^2$ for fully unscreened objects and $0$ for fully screened objects. For partial screening $\delta G$ may take any value between these limits. We show in the following section how this behaviour is manifest in cross-correlation functions when the galaxy subpopulations have different degrees of screening. In this case we will label $\delta G$ with a subscript to indicate which type of galaxy it refers to.

\section{Correlation Functions under Screened Fifth Forces}
\label{sec:corr}

The number density of galaxies traces the underlying density field in the universe on the largest scales. 
The overdensity in the number of galaxies at position $\textbf{x}$ is defined by $\Delta(\textbf{x}) \equiv \frac{N(\textbf{x}) - \bar{N}}{\bar{N}}$, where $N$ is the galaxy number density and $\bar{N}$ is the mean number density overall.
These overdensities contain a wealth of information about both the initial conditions of the universe and the distribution and properties of matter on cosmological scales \cite{PhysRevD.84.063505}. 
The derivation of the main effects contributing to galaxy overdensities at linear order, including observational effects, can be found in \cite{PhysRevD.80.083514, Bonvin:2013ogt, PhysRevD.84.063505, PhysRevD.84.043516}. We summarise them here. 
At a given redshift $z$, the overdensity of galaxies at an angular position $\hat{\textbf{n}}$ on the sky is given by 
\begin{eqnarray}
	& & \Delta(z, \hat{\textbf{n}}) = \Delta^{st}(z, \hat{\textbf{n}}) + \Delta^{rel}(z, \hat{\textbf{n}}) + \Delta^{lens}(z, \hat{\textbf{n}}) + \Delta^{AP}(z, \hat{\textbf{n}}) \label{start}\nonumber \\
	& & \Delta^{st}(z, \hat{\textbf{n}}) = b \delta(z, \hat{\textbf{n}}) - \mathcal{H}^{-1} \partial_r( \textbf{v} \cdot \hat{\textbf{n}}) \label{dst} \\
	& & \Delta^{rel}(z, \hat{\textbf{n}}) = \mathcal{H}^{-1} \partial_r \Psi + \mathcal{H}^{-1} \dot{\textbf{v}} \cdot \hat{\textbf{n}} \label{drel}\\
	& & - \left[ \frac{\mathcal{\dot{H}}}{\mathcal{H}^2} + \frac{2}{r\mathcal{H}} - 1 + 5 s \left( 1 - \frac{1}{r\mathcal{H}} \right) \right] \textbf{v} \cdot \hat{\textbf{n}}  \\
	& & \Delta^{lens}(z, \hat{\textbf{n}}) = (5s - 2) \int^r_0 \ dr' r' \left( \frac{r - r'}{2r} \right) \nabla^2_{\perp} (\Phi + \Psi)  \\
	& & \Delta^{AP}(z, \hat{\textbf{n}}) = \left( \partial_r - \partial_\eta \right) \left[ \Delta^{st} + \Delta^{rel} + \Delta^{lens} \right] \frac{dr (z, \hat{\textbf{n}})}{\partial \vec{\Theta}} \delta \vec{\Theta}.\nonumber  \\ \label{full_terms}
\end{eqnarray}
 $\vec{\Theta}$ is the cosmological parameter vector and $b$ is the linear bias between the galaxy density and the dark matter density $\delta$: $\Delta(\textbf{x}) = b \delta(\textbf{x})$. $\mathcal{H}$ is the conformal Hubble parameter and $\Psi$ and $\Phi$ are the weak-field metric potentials in the conformal Newtonian gauge:
\begin{equation}
	ds^2 = a^2(\eta) \left[ -\left( 1 + 2 \Psi \right) d\eta^2 + \left(1 - 2 \Phi \right) dx^i dx^j \delta_{ij} \right].
\end{equation}
$s$ describes the magnification bias that derives from the slope of the luminosity function:
 \begin{equation}
 	s \equiv \frac{2}{5} \left( \int df \ \epsilon(f) N_0(f) \right)^{-1} \int df \frac{d\epsilon}{df} \ f \ N_0(f).
\end{equation}
$N_0(f) df$ is the number density of sources with flux $f \pm \frac{df}{2}$ and $\epsilon(f)$ is the detection efficiency of those sources. 
$r$ is the comoving radial coordinate in the direction $\hat{\textbf{n}}$. 

The terms have been separated according to the physical effects that they embody. 
$\Delta^{st}$ contains the standard terms that relate the galaxy overdensity to the dark matter overdensity and the anisotropy caused by redshift space distortions.
This term is always accounted for in CF analyses.
$\Delta^{rel}$ contains the relativistic contributions such as the Doppler and integrated Sachs-Wolfe effects. This is the term which modified gravity effects alter: the acceleration terms are sourced by the Poisson equation and are therefore affected by the presence of a fifth force.
$\Delta^{lens}$ derives from the conversion of observed solid angle to physical solid angle given lensing along the line of sight. The final term describes the Alcock-Paczinski effect. We will only be interested here in the standard and relativistic terms, as it is their correlation that gives rise to the parity-breaking signal.

The expressions for the overdensities in Eqs. (\ref{start}--\ref{full_terms}) are the same in all metric theories of gravity where photons travel along null geodesics. This includes all theories conformally identical to GR, which includes most screened theories. Galaxies on the other hand are non-relativistic tracers of timelike geodesics, and are therefore directly affected by fifth forces. Their motion is governed by the Euler equation, which, in a perturbed FRW background can be written as 
\begin{equation}
	\dot{\textbf{v}} \cdot \hat{\textbf{n}} + \mathcal{H} \textbf{v} \cdot \hat{\textbf{n}} + \partial_r \Psi = 0.
\end{equation}
{Here we are using the conformal Hubble parameter $\mathcal{H}$. }
This can be used in Eq. \ref{drel} to give
\begin{equation}
	\Delta^{rel}(z,\hat{\textbf{n}}) = - \left[ \frac{\dot{\mathcal{H}}}{\mathcal{H}^2} + \frac{2}{r \mathcal{H}} + 5 s \left( 1 - \frac{1}{r \mathcal{H}} \right) \right] \textbf{v} \cdot \hat{\textbf{n}}.
\end{equation}
We see that the gravitational effects in the Euler equation cancel some of the terms, which is a manifestation of the equivalence principle.
As noted in \cite{Bonvin:2018ckp}, this is no longer true in the presence of a fifth force. According to Eq.~\ref{eq:F5} a long-range fifth force behaves identically to Newtonian gravity, enabling us to capture its effect with the transformation
\begin{equation}
	\partial_r\Psi \rightarrow (1 + \delta G) \: \partial_r \Psi, \label{FF}
\end{equation}
which describes a fractional increase in the strength of gravity by an amount $\delta G$. As discussed in Sec.~\ref{sec:screening}, the logic of screening implies that this is different for different galaxies as a function of their mass distributions and environments. 
{We make the simplifying assumption that $\delta G$ is a constant for each galaxy population, thereby ignoring the effect of partial screening. As we consider one population to be fully screened and the other fully unscreened, this provides an upper bound on the magnitude of their asymmetric cross-correlation. The simple modification of Eq.~\ref{FF} provides a clear intuitive picture of the physical origin of fifth force effects in the correlation function, and will also show clearly why these generate an octopole in the presence of screening, a key result of our paper. Deriving the exact degree of screening of each object would require solving the equation of motion of the scalar field numerically in the presence of a given density field (e.g.~\cite{Shao}), which is beyond the scope of this work.}

\begin{figure}[]
\begin{center}
\includegraphics[scale = 0.31]{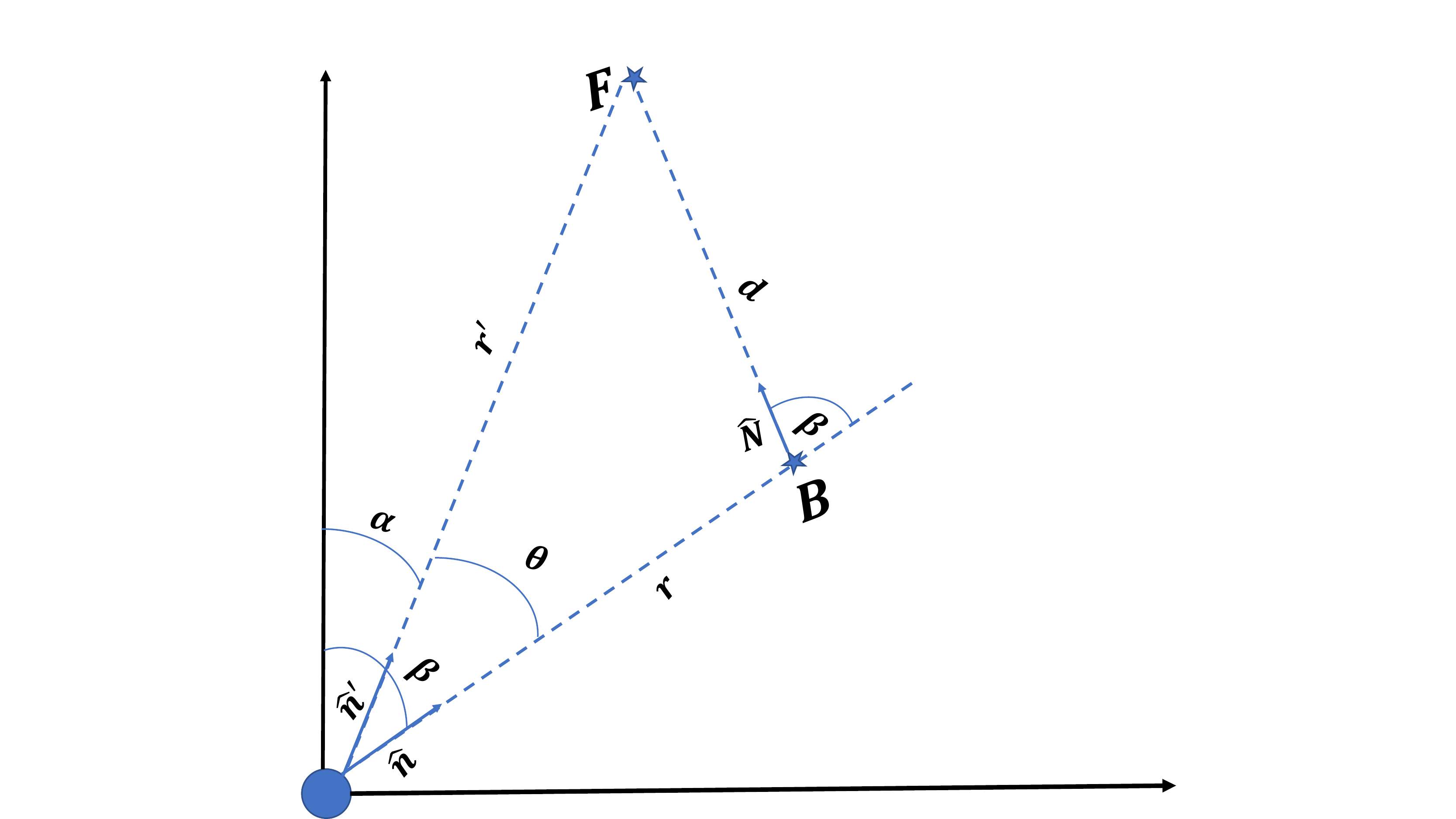}
\caption{Geometry of the CF. The observer is at the origin. $B$ and $F$ denote ``bright'' and ``faint'' galaxies; we are interested in their cross-correlation.}
\label{geometry}
\end{center}
\end{figure}

With this modification, the new expression for the relativistic part of the overdensity $\Delta^{rel(\mathcal{F})}$ is the sum of the usual relativistic term in Eq. (\ref{drel}) and an additional term $\Delta^{\mathcal{F}}$ due to the fifth force:
\begin{eqnarray}
	\Delta^{rel(\mathcal{F})}(z, \hat{\textbf{n}}) & \equiv & \Delta^{rel}(z, \hat{\textbf{n}}) + \Delta^{\mathcal{F}}(z, \hat{\textbf{n}}) \nonumber \\
	\Delta^{\mathcal{F}}(z, \hat{\textbf{n}})  & = & \zeta \left[ \frac{ \dot{\textbf{v}} \cdot \hat{\textbf{n}}}{\mathcal{H}} + \textbf{v} \cdot \hat{\textbf{n}} \right] \label{drel_screen}
\end{eqnarray}
where $\zeta \equiv \left( \frac{\delta G}{1 + \delta G} \right)$. This replaces $\Delta^{rel}$ in the total expression for $\Delta$.\footnote{Our parametrisation of modified gravity is related to that of \cite{Bonvin:2018ckp} by $\Gamma = \delta G$, $\Theta = 0$; thus, although somewhat more general, their model does not account for screening as it effectively assigns the same $\delta G$ to all galaxies.} $\zeta = 0$ in the case of complete screening (fifth force fully suppressed) and $\zeta = \zeta_\text{max}$ for objects that are fully unscreened. $\zeta_\text{max}$ is set by the coupling coefficient of the scalar field to matter; for example it is 1/4 in $f(R)$ where $\delta G = 1/3$. As $\zeta$ for a given galaxy depends on its mass distribution and gravitational environment via its degree of screening, we assign the two galaxy populations (which we denote ``bright'', $B$, and ``faint'', $F$, as in Sec.~\ref{sec:intro}) different average values $\zeta_B$ and $\zeta_F$.
We can now compute the parity-breaking CF.
The cross-correlation between the $B$ and $F$ populations is given by $\langle \Delta_B(\textbf{x}_1) \Delta_F(\textbf{x}_2) \rangle$, which we write, with the geometry of Fig.~\ref{geometry}, as
\begin{equation}
	\xi(z,z', \theta)_{BF}= \langle \Delta_B(z, \hat{\textbf{n}}) \Delta_F(z', \hat{\textbf{n}}') \rangle.
\end{equation}
Due to the assumption of statistical isotropy this depends only on the angle $\theta$ between the galaxies as they are projected on the sky. Furthermore, it is important to remember that $z, z'$ and $\theta$ are \emph{observed} redshifts and angular sizes. Converting these to \emph{physical} quantities depends on the background cosmology, although to linear order the corrections from this are already accounted for in the expressions for the overdensities.

In principle there is a CF for each term in $\Delta$, although we are only interested here in the relativistic part $\xi^{rel(\mathcal{F})}$ which is sensitive to the fifth force
\begin{eqnarray}
	\xi^{rel(\mathcal{F})}(z, z', \hat{\textbf{n}}) & = & \langle \Delta^{st}_B(z,\hat{\textbf{n}}) \Delta^{rel(\mathcal{F})}_F  (z', \hat{\textbf{n}}') \rangle \nonumber \\
	& + &  \langle \Delta^{st}_F(z', \hat{\textbf{n}}') \Delta^{rel(\mathcal{F})}_B (z, \hat{\textbf{n}}) \rangle .
\end{eqnarray}
The individual CFs are calculated in Appendix \ref{corr_funcs}; here we show the final result, expanded to leading order in $d/r \ll 1$:
\begin{widetext}
\begin{eqnarray}
	& & \xi^{rel(\mathcal{F})}(r,d, \beta) = \xi^{rel}(r,d, \beta) + \xi^{(\mathcal{F})}(r,d, \beta) \nonumber \\
	& & \xi^{rel}(r,d, \beta) = \frac{2 A_s}{9 \pi^2 \Omega_m^2} \frac{\mathcal{H} D^2 f}{\mathcal{H}_0}\left\{ P_1(\cos \beta) \nu_1(d) \left[ (b_B - b_F) \left( \frac{\dot{\mathcal{H}}}{\mathcal{H}^2} + \frac{2}{r \mathcal{H}} \right) - \left( 1 - \frac{1}{r \mathcal{H}} \right) \left( 5(s_Bb_F - s_F b_B) + 3f(s_B - s_F) \right) \right]  \right. \nonumber \\
	& &\left. + 2 P_3(\cos \beta) \nu_3(d) \left( 1 - \frac{1}{r \mathcal{H}} \right) f(s_B-s_F) \right\} \nonumber \\
	& & \xi^{(\mathcal{F})}(r,d, \beta) = \frac{2 A_s M}{9 \pi^2 \Omega^2_m} \left\{ P_1(\cos \beta) \nu_1(d) \left[ (b_F \zeta_B - b_B \zeta_F) - \frac{3}{5} f( \zeta_B - \zeta_F) \right] + P_3(\cos \beta) \nu_3(d) \left[ \frac{2f}{5} (\zeta_B - \zeta_F) \right] \right\}. \label{eq:final}
\end{eqnarray}
\end{widetext}
We have defined $f \equiv \frac{d \ln D}{d \ln a}$, where $D$ is the linear growth factor. 
{Throughout our calculations we have assumed these are scale-independent, following \cite{Bonvin:2018ckp}\footnote{{In principle this can generalised and the growth factor made scale dependent. We plan on exploring this for more generic modified gravity theories in the modified Boltzmann code HiClass \cite{hiclass} in future works. We take the first steps in this direction by writing our screening model in Horndeski form in Appendix~\ref{horndeski}} }.}
$P_n$ is the Legendre polynomials of order $n$, $b_B$ and $b_F$ are the biases for the bright and faint galaxies respectively, and 
\begin{eqnarray}
	& & \nu_\ell(d) \equiv \int d \ln k \ (k \eta)^{n_s-1} \left( \frac{k}{\mathcal{H}_0} \right)^3 j_\ell(kd) T^2(k), \nonumber \\
	& & M(a) \equiv \frac{D^2}{\mathcal{H}_0 \mathcal{H}} \left( \dot{\mathcal{H}} f + \mathcal{H} \dot{f} + f^2 \mathcal{H}^2 + f \mathcal{H}^2\right). \label{nu_stuff}
\end{eqnarray}
We have dropped all terms in $\xi^{rel}$ that involve the correlation of two relativistic terms ($\langle \Delta^{rel} \Delta^{rel} \rangle$) as these are suppressed by factors of $\frac{\mathcal{H}}{k}$.

As anticipated in Sec.~\ref{sec:intro}, besides modifying the cosmological background the fifth force affects the CF in two distinct ways. The first is the reintroduction of redshift terms due to the difference induced between null and timelike geodesics (first term in $\xi^{(\mathcal{F})}$). This would also be present under a universally-coupled (i.e. non-screened) fifth force and is the type of modified gravity considered by \cite{Bonvin:2018ckp}. The second is the relative effect on the $B$ and $F$ populations due to their different sensitivities to fifth forces under a screening mechanism. This is shown by the remaining terms in $\xi^{(\mathcal{F})}$ which are proportional to $\zeta_B - \zeta_F$. We see that the fifth force introduces both a dipole (the term proportional to to $P_1\nu_1$) and an octopole (the term proportional to $P_3 \nu_3$). While the dipole is increased by both screened and universal fifth-force terms, in the absence of screening the octopole is only present if $s_B \ne s_F$. Thus, as discussed in more detail below, the octopole may provide a particularly clean probe of screening.

\section{Correlation function parameter space}
\label{sec:parameters}

The CF of Eq.~\eqref{eq:final} depends on several parameters of standard $\Lambda$CDM, as well as those of modified gravity. The aim of this section is to explain and quantify the effects of these parameters. We begin by classifying them into two sets: 1) Those that affect the background cosmology and hence all CFs, and 2) Those that are specific to the parity-breaking CF and describe the environmental dependences of galaxy formation. We term the first set \emph{global} and the second set, whose members carry a $B$ or $F$ index, \emph{local}. The parameters are listed in Table \ref{fid_cosmo}, along with their fiducial values which we use throughout our analysis unless otherwise stated.

On a practical note, the global parameters take longer to evaluate as they affect background quantities such as the matter power spectrum: thus each point in parameter space corresponds to a run of a Boltzmann code. The local parameters are simply multiplicative factors in front of the functions of global parameters, making their parameter space in principle much quicker to explore.

\begin{center}
\begin{table}[h]
\begin{tabular}{ p{2cm}  p{2cm}}
\hline
\hline
$A_s$ & 2.1 $\times 10^{-9}$  \\
$h$ & 0.7 \\
$\Omega_bh^2$ & 0.0224  \\
$\Omega_{c}h^2$ & 0.112 \\
$k_{*}$ & 0.05 Mpc$^{-1}$ \\
$n_s $ & 0.96 \\
$f_{R0}$ & $0$ \\
\hline
\hline
$(b_B, b_F)_{z = 0}$ & (1.7, 0.84) \\
$(b_B^{f(R)} b_F^{f(R)})_{z = 0}$ & (1.64, 0.8) \\
$(s_B, s_F)$ & (0.1,  0) \\
$(\delta G_B, \delta G_F)$ & (0, 1/3) \\
\hline
\hline
\end{tabular}
\caption{Global (upper) and local (lower) parameters, along with their fiducial values. The bias values used in the $f(R)$ plots are different from GR and are denoted  by the $f(R)$ superscript index.}\label{fid_cosmo}
\end{table}
\end{center}

\subsection{Global parameters}

\begin{figure}
\begin{center}
\includegraphics[scale = 0.32]{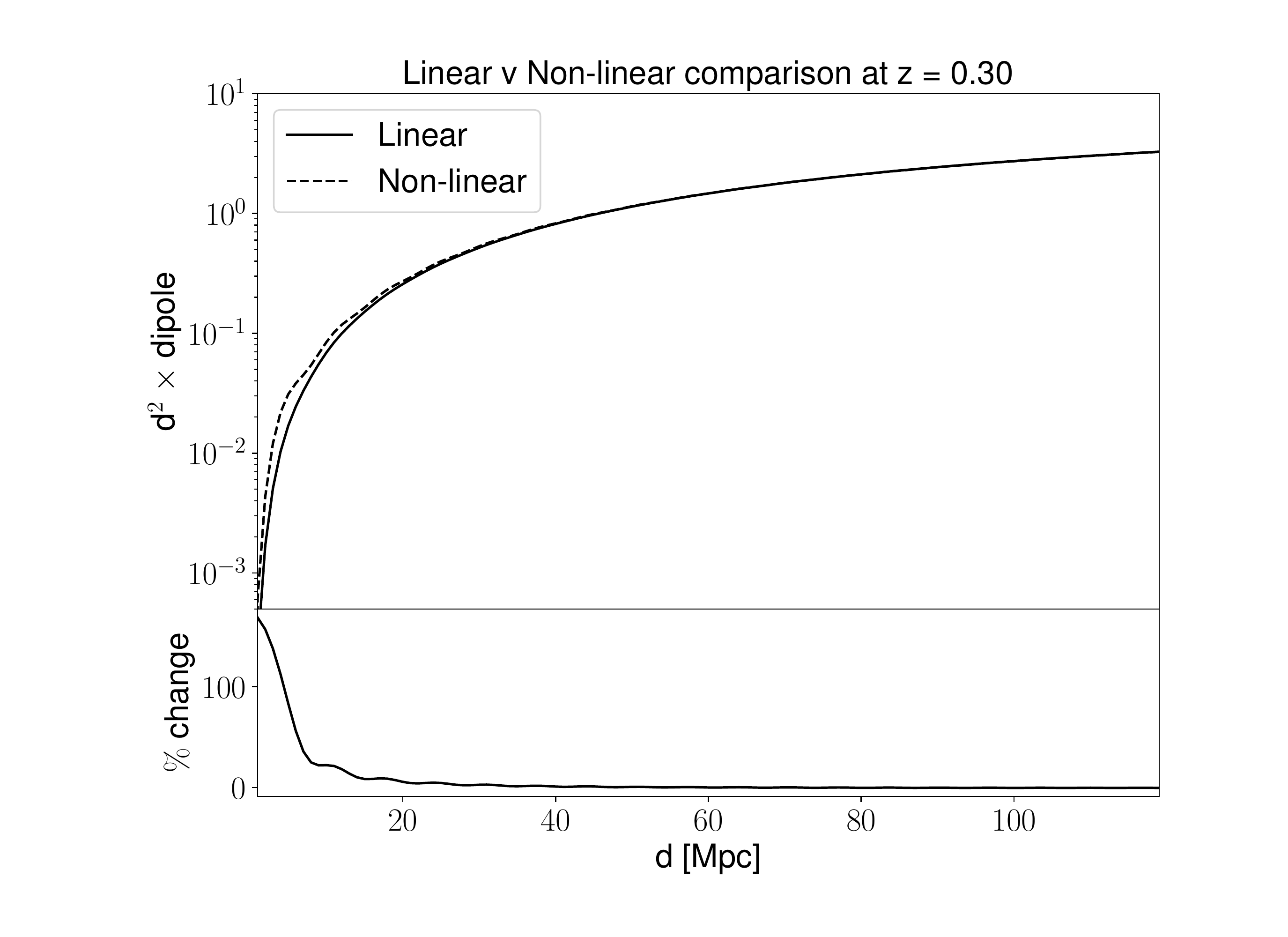}
\caption{The change to the $z=0.3$ $\Lambda$CDM dipole when a linear or non-linear power spectrum/transfer function (from Halofit) is used, relative to the linear case.}
\label{NL}
\end{center}
\end{figure}

There are three groups of global parameters. The first contains the standard $\Lambda$CDM parameters. As the effects of these on CFs have already been extensively studied \cite{Gaztanaga:2015jrs, Hall:2016bmm, Alam_mag, Giusarma:2017xmh, Alam:2017cja, Alam:2017izi}, we do not investigate them further here. The second set quantifies the effects of ``galaxy formation" physics, which generates non-linear corrections to the transfer function. We check the importance of this by using the Halofit fitting function \cite{Peacock:2000qk} to obtain the non-linear matter power spectrum, and show its effect on the dipole in Fig.~\ref{NL}. We see that on scales below $\sim$10 Mpc non-linearities lead to an $\mathcal{O}(1)$ modification, but the difference decays away rapidly on larger scales. From now on we will only use the transfer functions and power spectra with these non-linearities included.

The final set of global parameters describes the effect of modified gravity on the background perturbations, specifically the transfer function and power spectrum. To compute this effect we specialise to the case of Hu-Sawicki $f(R)$ \cite{Hu_Sawicki}, an archetypal and well-studied chameleon-screened theory known to be stable to instabilities, propagate gravitational waves at the speed of light and be capable of screening the Milky Way to pass local fifth-force tests. Although representative of the chameleon mechanism, Hu-Sawicki occupies only a small part of the full chameleon parameter space. A general chameleon model introduces three new degrees of freedom. At the level of the Lagrangian these are, for example, the $\{n, \Lambda, M\}$ of \cite{Burrage} Eq. 2.5 (see also Appendix~\ref{horndeski}). Phenomenologically they are the strength of the fifth force between unscreened objects (related to $\alpha$ in Eq.~\ref{f_Poisson}), the range of the fifth force (Compton wavelength of the scalar field) and the self-screening parameter $\chi$ (e.g. \cite{Burrage} eq. 3.2) that determines the threshold Newtonian potential at which screening kicks in. In $f(R)$ the coupling coefficient of the scalar field to matter is fixed at $\alpha = 1/\sqrt{6}$ ($\delta G = 1/3$) while the parameters $n$ and $\Lambda$ describing the field's potential are related.

In the Jordan frame, the Hu-Sawicki action is given by
\begin{equation}
S = \int d^4x \sqrt{-g} \left(\frac{M_\text{pl}^2}{2}(R+f(R)) + \mathcal{L}_m\right), \label{fR_HS}
\end{equation}
where
\begin{equation}\label{eq:fR}
f(R) = -m^2 \frac{c_1 (R/m^2)^k}{c_2(R/m^2)^k+1}
\end{equation}
and $\mathcal{L}_m$ is the matter Lagrangian. The mass scale is set by the average density of the Universe, $m^2 = \rho_m/(3 M_\text{pl}^2)$, and $c_1$, $c_2$ and $k$ are dimensionless free parameters. It is shown in \cite{Li:2011vk} that only $k$ and $\frac{c_1}{c_2^2}$ affect the matter power spectrum and thus the model only contains two relevant degrees of freedom. $k=1$ is a standard choice that we adopt here. In the Einstein frame this is a scalar--tensor theory in which the scalar field is $f_R \equiv df/dR$, the present value of which is also completely determined by $\frac{c_1}{c_2^2}$. The field at the cosmological background value of $R$, $f_{R0}$, determines the structure formation history of the Universe as well as the range and screening properties of the fifth force at a given epoch, and is effectively the theory's only degree of freedom. GR is recovered in the limit $f_{R0} \rightarrow 0$, and values in the range $\sim10^{-4}-10^{-6}$ have observable consequences in galaxy clustering, redshift space distortions, cluster abundance, intensity mapping and the matter bispectrum (see e.g. \cite{Arnalte-Mur:2016alq, li_2013, Bose:2014zba, Lombriser_rev} and references therein). For the Solar System to be screened requires $f_{R0} \lesssim 10^{-6}$. Smaller values may be probed by galaxy-scale tests \cite{Vikram, Vikram_RC, Cepheid, Desmond_2, Desmond_3}, which now rule out $f_{R0} > \text{few} \: \times 10^{-8}$ \cite{Desmond_1}.

For one of our galaxy types to self-screen and the other not, the screening parameter $\chi$ must be between their characteristic Newtonian potentials. This can be written in terms of the background scalar field value as $\chi \simeq 3/2 \: f_{R0}$. The halo masses calculated in Sec.~\ref{sec:local} imply $|\Phi_F| \simeq 1\times10^{-6}$ and $|\Phi_B| \simeq 6\times10^{-6}$ ($c\equiv1$). The galaxies may however be partly environmentally screened, increasing the required value of $\chi$.\footnote{To determine the screening properties of both galaxy populations one would ideally solve the equation of motion for the scalar field given the mass distribution around the galaxies, or at least a proxy for the field such as the Newtonian potential \cite{Desmond_maps}. However, as we are interested here in the general effects of a screened fifth force we leave this more detailed investigation for future work.} $10^{-6} \lesssim f_{R0} \lesssim 10^{-5}$ is therefore likely to separate the galaxies by screening properties, and also causes the scalar field to mediate an astrophysical-range fifth force \cite{Hu_Sawicki}. This is therefore the range that we consider.

\begin{figure}
\begin{center}
\includegraphics[scale = 0.33]{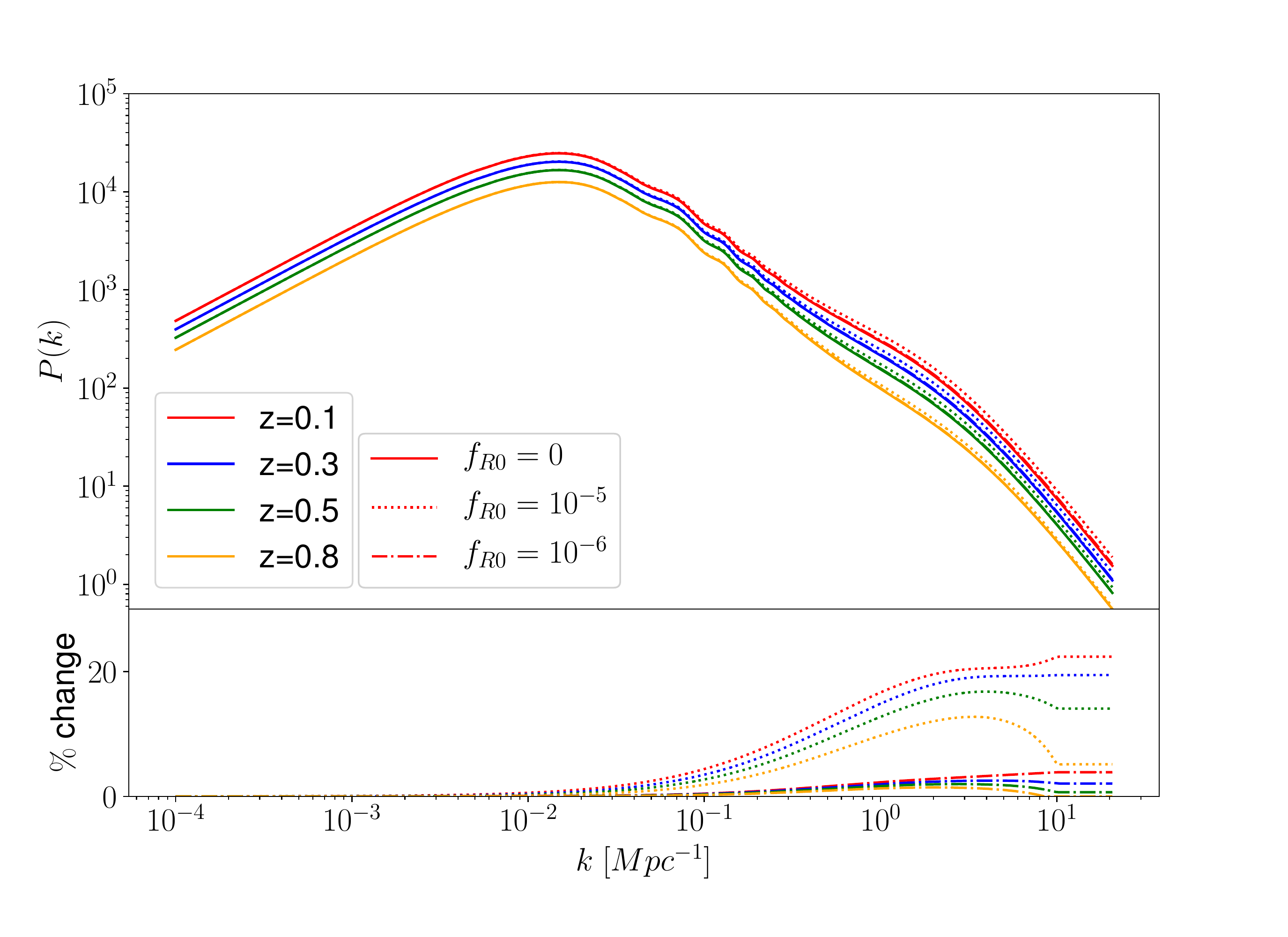}
\caption{The matter power spectrum evaluated in $\Lambda$CDM and $f(R)$ for a range of redshift and $f_{R0}$ values. The percentage difference is plotted with respect to the $\Lambda$CDM values with the fiducial parameters of Table \ref{fid_cosmo}.}
\label{mPk_fR}
\end{center}
\end{figure}

We compute the matter power spectrum using a modified version of CAMB calibrated with $N$-body simulations in the $k=1$ model \cite{Winther}, and plot this for different $f_{R0}$ values at various redshifts in Fig.~\ref{mPk_fR}. We see that the power spectrum changes by $\sim 20 \%$ on scales smaller than $\sim 1$ Mpc for $f_{R0}  = 10^{-5}$, while for $f_{R0} = 10^{-6}$ the change is only a few percent. This is propagated into the CF in Sec.~\ref{sec:results}.

\subsection{Local parameters}
\label{sec:local}

We consider three local parameters: galaxy bias $b$, magnification bias $s$ and fifth-force sensitivity $\delta G$. The first two are present in $\Lambda$CDM and give rise to the standard parity-breaking CF, while the latter is the specific focus of our study. We describe our choices for these parameters below.

\begin{itemize}

\item $(b_B, b_F)$: As our fiducial case we take the $B$ galaxies to be luminous red galaxies and the $F$ galaxies to be emission line galaxies, with biases {in $\Lambda$CDM} of 1.7 and 0.84 respectively at $z=0$ \cite{DESI}. These are typical populations that will be measured by the forthcoming DESI survey, which will provide the next significant improvement in measurement of the parity-breaking CF.\footnote{Forecasts for testing gravity with the parity-breaking CF for non-screened theories can be found in \cite{Bonvin:2018ckp}} Under the Sheth--Tormen model \cite{Sheth_Tormen} these biases correspond to halo masses $\sim$10$^{13} \: h^{-1} M_\odot$ and $\sim$10$^{14} \: h^{-1} M_\odot$ respectively. When showing results for $z>0$ we model the redshift dependence of the bias as \cite{Tegmark:1998wm, Fry:1996fg, Nusser:1993sx, Bonvin:2013ogt}
\begin{equation}
	b_{B / F}(z) = 1 + (b_{B / F}(z = 0) -1) \frac{D(z = 0)}{D(z)} \label{bias}
\end{equation}
with $D$ the regular growth factor.
We check how the dipole is affected by a change in the bias and as a function of redshift in Fig.~\ref{dip_local}. {To account for the fact that the bias is reduced in $f(R)$ due to the action of the fifth force (e.g. \cite{Schmidt:2008tn}), we take $b_B^{f(R)}(z=0) = 1.62$, $b_F^{f(R)}(z=0) = 0.8$ when we study $f_{R0}=10^{-5}$.\footnote{{Taken from fig. 14 of \cite{Arnold:2018nmv} for galaxies of mass $10^{13} M_\odot/h$ and $10^{14} M_\odot/h$ for $b_F(z=0)$ and $b_B(z=0)$. We take the mean value from both box sizes and thus change the bias by $\sim$ -4$\%$ for both types of galaxies.}} We also use the growth function in $f(R)$  theory to compute the change in bias at different redshifts. This leads to a $\mathcal{O}(10\%)$ change in the bias at $z = 1$. These bias values are used for the $f(R)$ cases of Figs.~\ref{dip_oct} and \ref{pchange}.}

\item $(s_B, s_F)$: If the magnification bias is the same for both galaxy populations then the octopole in $\Lambda$CDM is zero. Under this common assumption the octopole is a unique signature of \emph{screened} fifth forces: as shown in Eq.~\ref{eq:final}, even modified gravity without screening does not produce it. However, in order to gauge the relative importance of magnification bias and fifth force, we consider a plausible value of $s_B - s_F = 0.1$ \cite{Alam_mag, Durrer_1, Durrer_2}.

\item $(\delta G_B, \delta G_F)$: There are various considerations for setting the values of $\delta G_B$ and $\delta G_F$. $\delta G_B$ should be no larger than $\delta G_F$ because brighter galaxies should be more massive (and occupy denser environments), and hence more screened. Our requirement that $|\Phi_F| \lesssim \chi \lesssim |\Phi_B|$, and assumption of $f(R)$ in the background, implies $\delta G_B = 0$, $\delta G_F = 1/3$ as our fiducial choice. It is worth noting, however, that one can construct theories in which the change to the transfer function and growth rate is small but the unscreened $\delta G$ is $\mathcal{O}(1)$ or larger, in which case the predicted signal simply scales the $\Lambda$CDM result linearly with $\delta G$. The parity-breaking CF would provide maximal sensitivity to screening \textit{per se} in such a scenario, due to the insignificance of modified gravity in the background.

\end{itemize}

\section{Results}
\label{sec:results}

In this section we calculate the dipole and octopole numerically in our model. We use the Boltzmann code CAMB \cite{Lewis:1999bs} to compute the transfer functions in $\Lambda$CDM, along with the modification presented in \cite{Winther} for $f(R)$. Throughout the computations we use the fiducial parameters presented in Table \ref{fid_cosmo} and a range of redshifts, $ z \in \{0.1, 0.3, 0.5, 0.8\}$. 

\begin{figure*}
\centering
\begin{subfigure}{.5\textwidth}
  \centering
  \includegraphics[width=\linewidth]{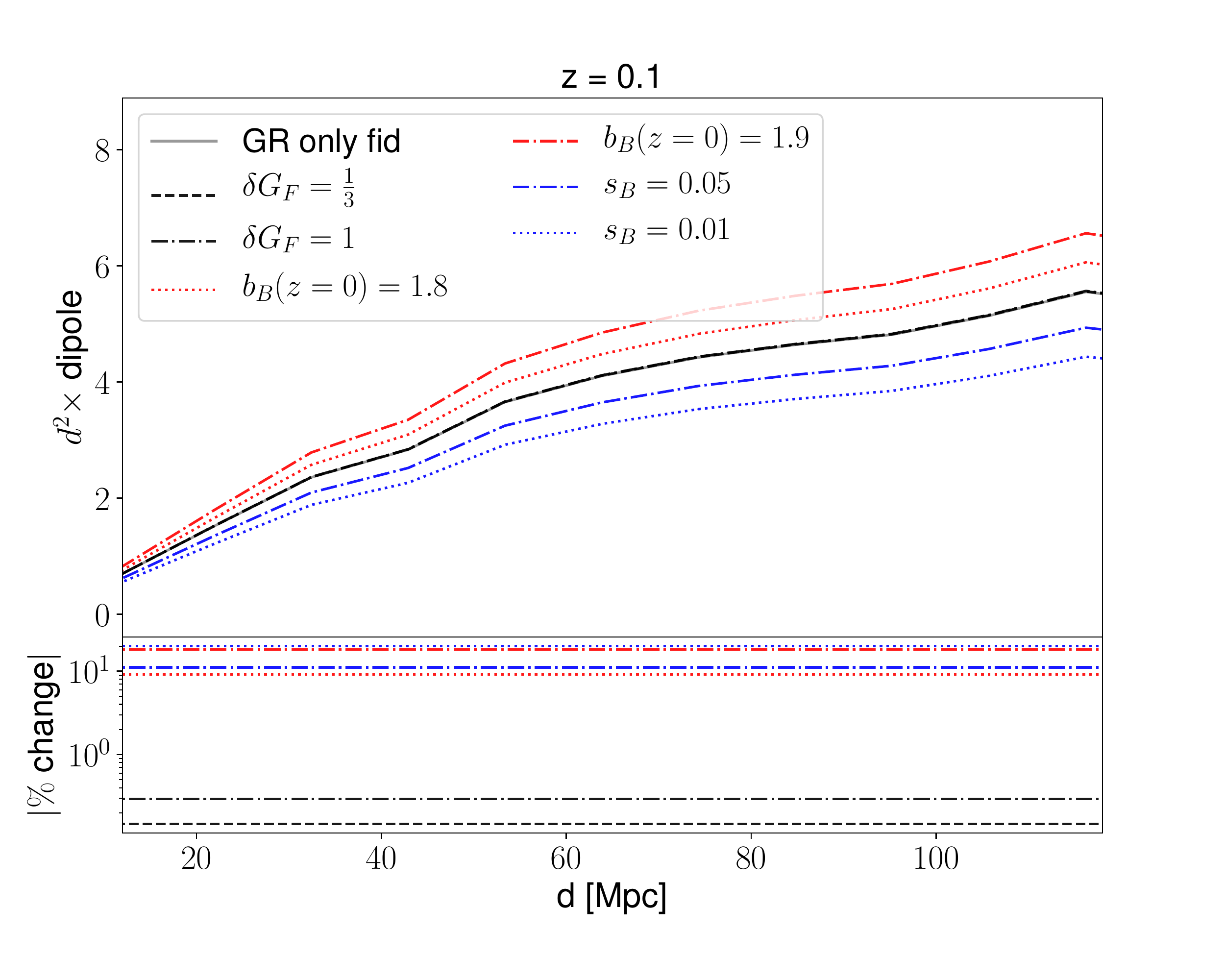}
  \caption{}
  \label{fig:sub1}
\end{subfigure}%
\begin{subfigure}{.5\textwidth}
  \centering
  \includegraphics[width=\linewidth]{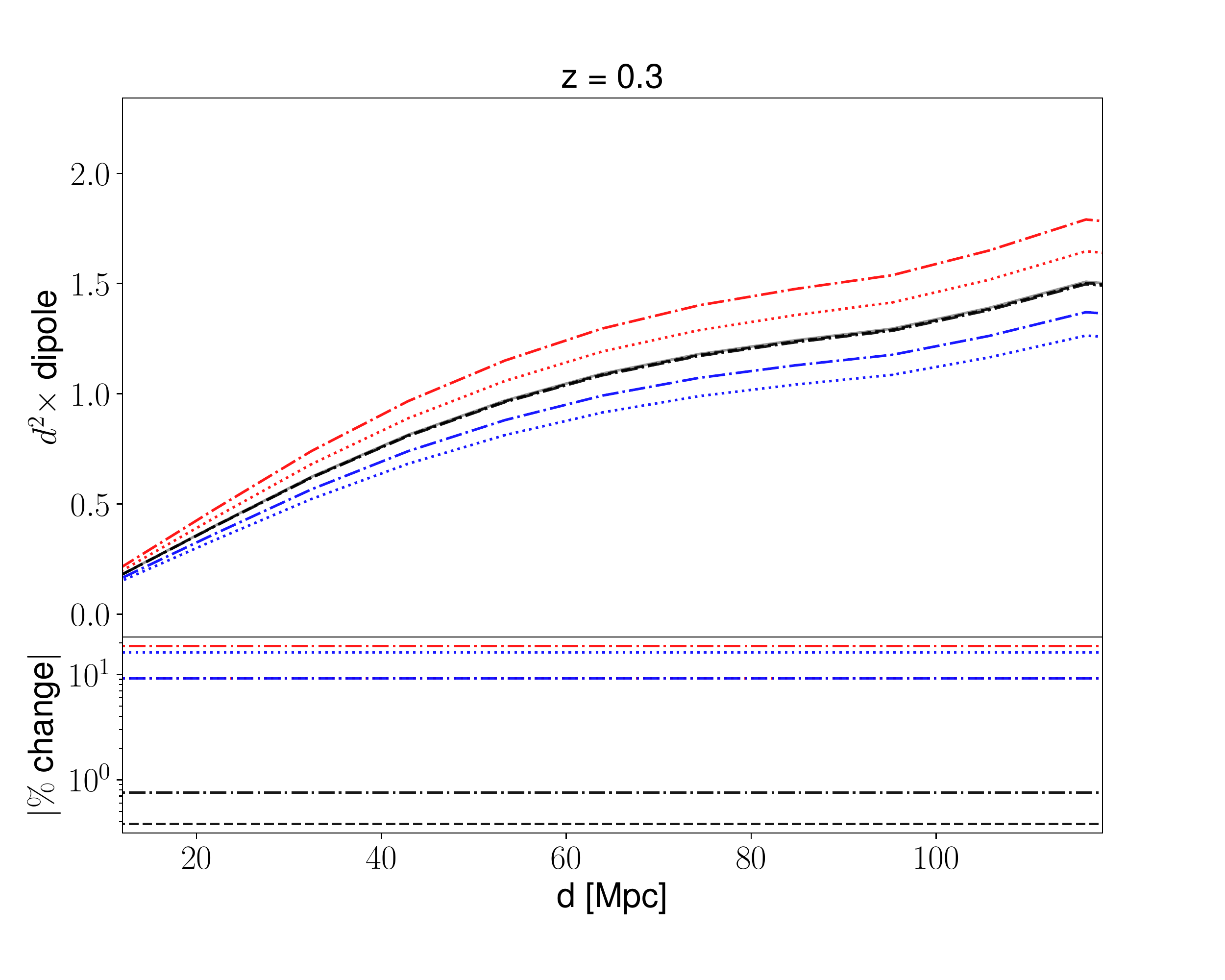}
  \caption{}
  \label{fig:sub2}
\end{subfigure}

\begin{subfigure}{.5\textwidth}
  \centering
  \includegraphics[width=\linewidth]{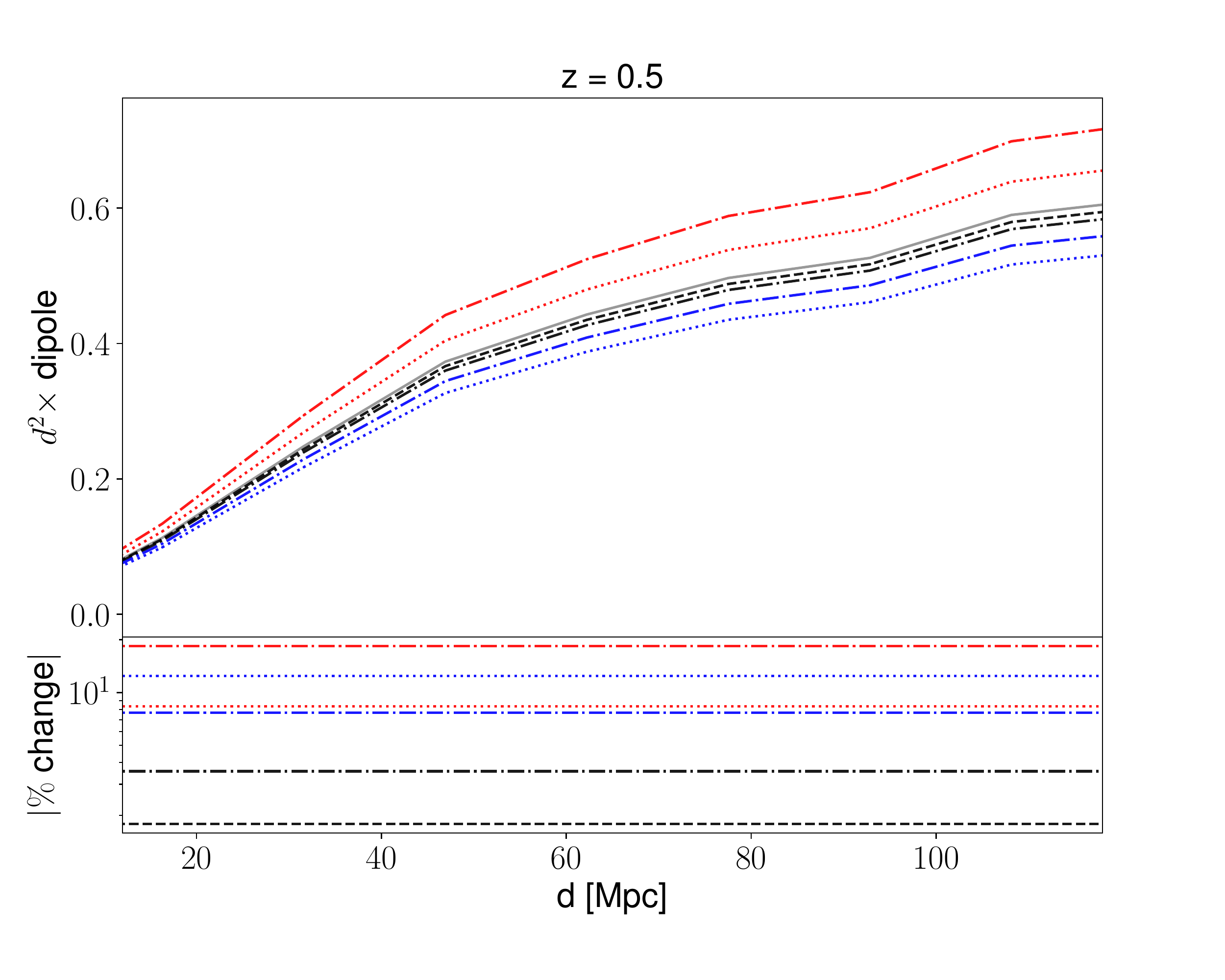}
  \caption{}
  \label{fig:sub1}
\end{subfigure}%
\begin{subfigure}{.5\textwidth}
  \centering
  \includegraphics[width=\linewidth]{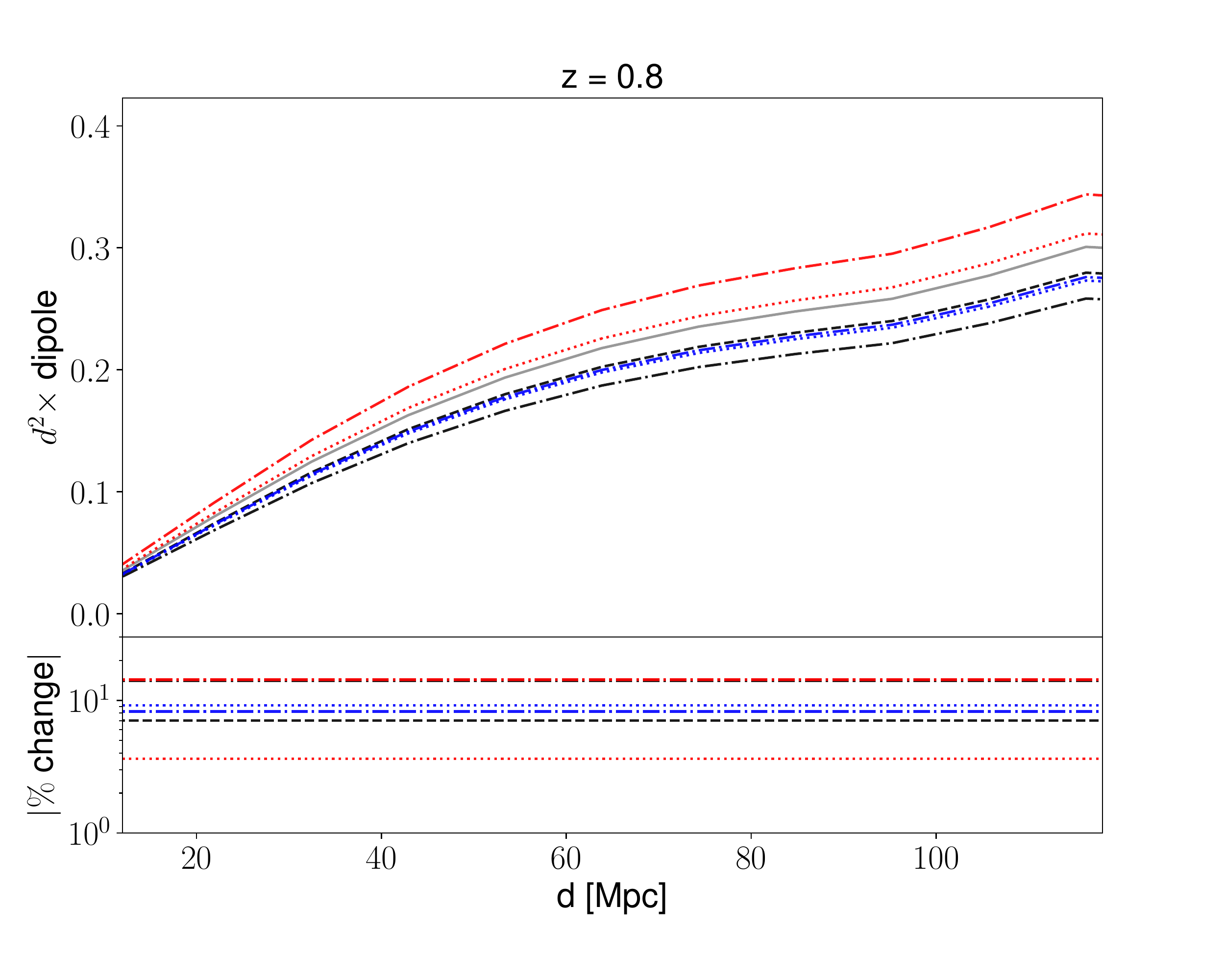}
  \caption{}
  \label{fig:sub2}
\end{subfigure}
\caption{CF dipole for various values of local parameters in a $\Lambda$CDM background, as listed in the legend of subfigure (a) and the same in all cases. Changes are defined with respect to the fiducial values listed in Table~\ref{fid_cosmo}. Each subfigure corresponds to a particular redshift, as indicated. The lower panels show the magnitudes of the percentage changes with respect to the fiducial $\Lambda$CDM dipole (grey line).}
\label{dip_local}
\end{figure*}

First we check the sensitivity of the dipole to local parameters. Fig.~\ref{dip_local} shows the dipole at various redshifts for a range of galaxy bias, magnification bias and $\delta G$ values. We assume throughout that the $B$ galaxies are completely screened while varying the screening felt by the $F$ galaxies; conversely, when investigating bias we fix $b_F$ and $s_F$. We see that $\delta G_F = 1/3$, which corresponds to complete unscreening in $f(R)$, changes the dipole by $\sim\mbox{few} \times 0.1\%$ for $z = 0.1, 0.3$, whereas at $z = 0.5, 0.8$ the change is $\sim \mbox{few} \times 1 \% - 10\%$. A $\sim 10\%$ change in the galaxy bias changes the dipole by the same amount. Decreasing the magnification bias by a factor of (2, 5) decreases the dipole by $\sim (10, 15)\%$ at all redshifts except $z = 0.8$ where the effect is slightly smaller. Interestingly, at $z = 0.8$ the percentage change due to screening is roughly the same as the effect of varying the bias in the range we consider. This shows that to able to detect the modification due to screening it will be necessary to know or model the galaxy bias to $\sim0.1-1\%$ for $z = 0.1, 0.3$, but only $\sim 5, 10 \%$ for $z = 0.5, 0.8$. It is also worth noting that the galaxy bias is typically measured in combination with $\sigma_8$ in clustering analyses, while breaking the degeneracy with $\sigma_8$ requires information from weak lensing. The magnification bias would need to be known to a within a factor $\sim$10 to constrain screening parameters at low redshift, while at higher redshift it must be known to within a factor $\sim$2. 

\begin{figure*}
\centering
\begin{subfigure}{.5\textwidth}
  \centering
  \includegraphics[width=\linewidth]{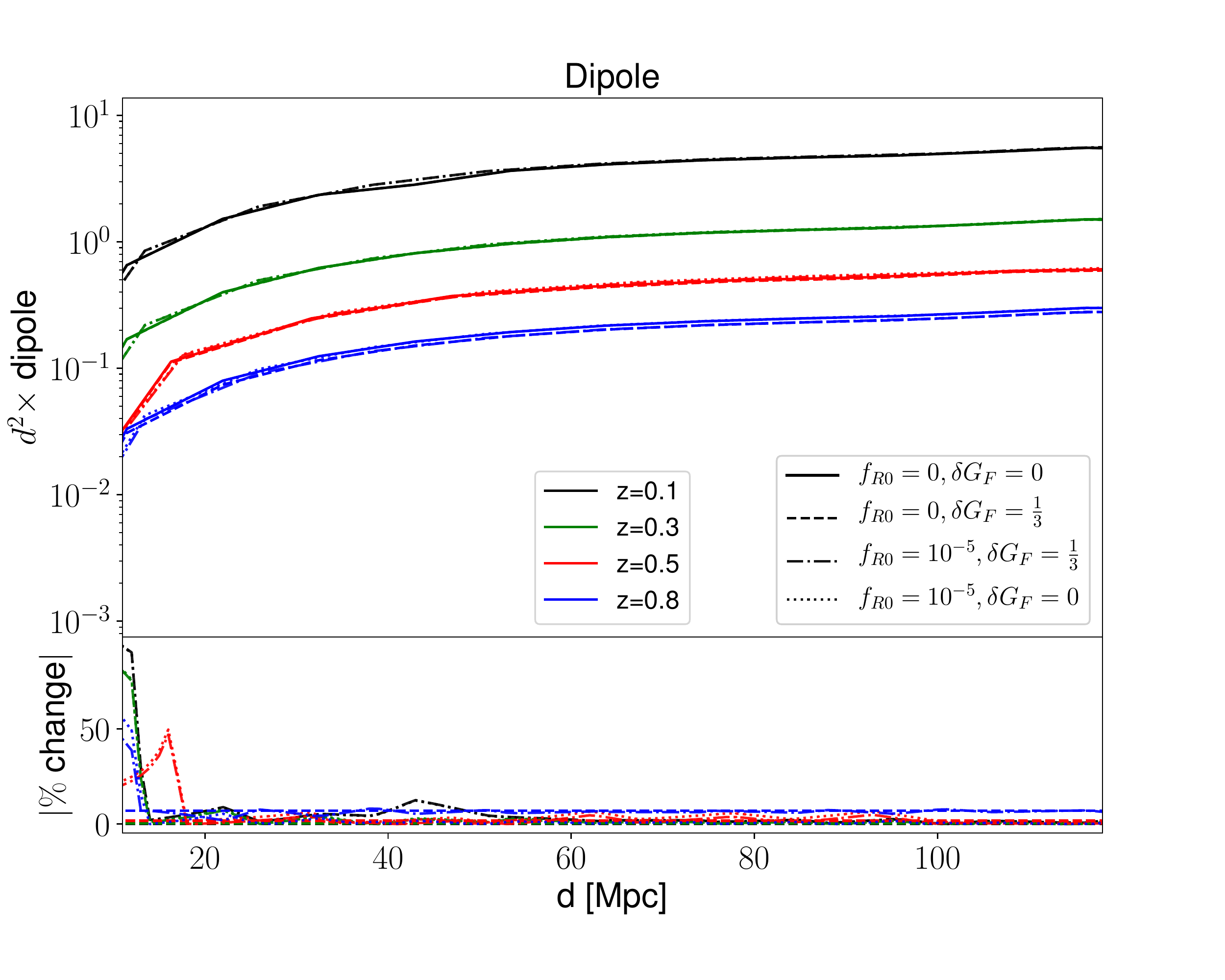}
  \caption{The dipole for a range of models involving local screening and/or modification to the cosmological background. The solid line is the fiducial $\Lambda$CDM dipole at each redshift, which the percentage differences are defined relative to.}
  \label{global_dip}
\end{subfigure}%
\begin{subfigure}{.5\textwidth}
  \centering
  \vspace{3mm}\includegraphics[width=\linewidth]{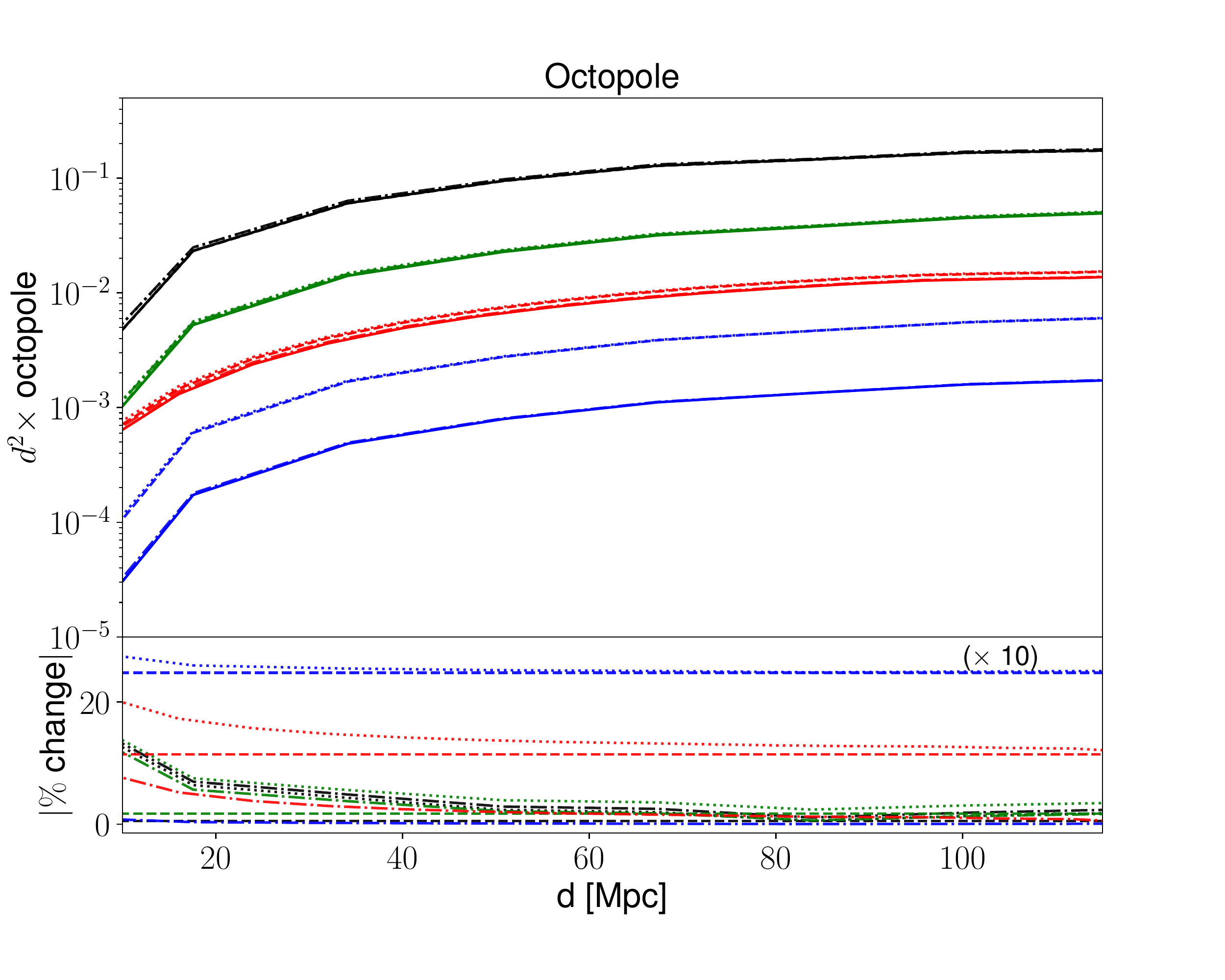}
   \caption{The absolute value of the octopole for the same models as in Fig.~\ref{global_dip}. The change in the octopole for different local parameter values can be read off from Eq.~\eqref{eq:final}. We scale the blue line down by a factor of 10 in the percentage change in order for it to fit on the plot: the true value is $\sim$200\%.}
  \label{global_oct}
\end{subfigure}
\caption{Dipole and octopole components of the parity-breaking CF in $\Lambda$CDM and $f(R)$ for $0.1 < z < 0.8$.}
\label{dip_oct}
\end{figure*}

Next we show in Fig.~\ref{global_dip} the dipole for $f_{R0}=10^{-5}$, which generates a cosmological-range fifth force. We see that the dipole changes by $\mathcal{O}(1)$ on scales $\lesssim 10$ Mpc, but less at higher redshift. For $z = 0.5$ and below the change is $\mathcal{O}(0.1-1\%)$ on scales $\gtrsim 20$ Mpc. Thus the effect from the change in background in $f(R)$ theories appears to dominate the effect of realistic screening values by an order of magnitude for $z \lesssim 0.5$, while at higher $z$ the screening effect can be twice as important as the change in background. Thus the most promising regime in which to search for signs of screening in the dipole is $z > 0.5$, while at lower redshift one should hope instead to detect the change due to the effect of modified gravity on the growth rate and transfer function.

In Fig.~\ref{global_oct} the octopoles for both $\Lambda$CDM (with and without screening) and $f(R)$ background with $f_{R0} = 10^{-5}$ are computed. We see that in the case of $f_{R0} = 0$ (i.e. a local screened fifth force in a $\Lambda$CDM background), the octopole changes by few $\times 1 \%$ for $z = 0.1, 0.3$, whereas for $z = (0.5, 0.8)$ it changes by $\sim (10\%, 25\%)$. This always dominates over the effect of background modification only (no local screening). It is worth bearing in mind that in $\Lambda$CDM (and non-screened modified gravity), an octopole only arises due to the difference in magnification bias between the two galaxy populations, which is typically assumed to be zero. In this case \emph{any} octopole would be a sign of screening. The results for any local parameter choices can be readily constructed as they simply multiply the curves in Fig.~\ref{global_oct}.

\begin{figure*}
\centering
\begin{subfigure}{.5\textwidth}
  \centering
  \includegraphics[width=\linewidth]{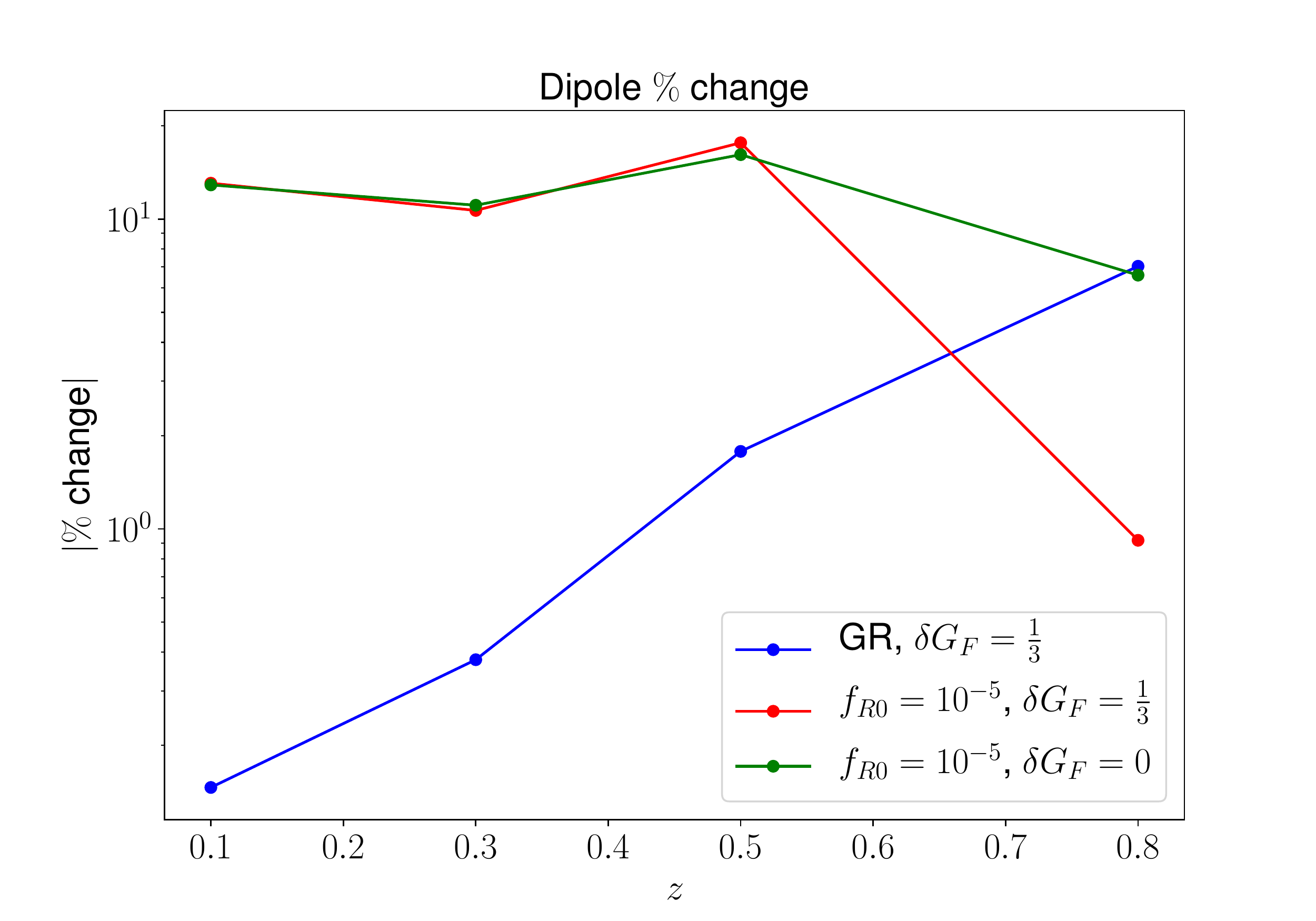}
  \caption{}
  \label{pc_dip}
\end{subfigure}%
\begin{subfigure}{.5\textwidth}
  \centering
  \includegraphics[width=\linewidth]{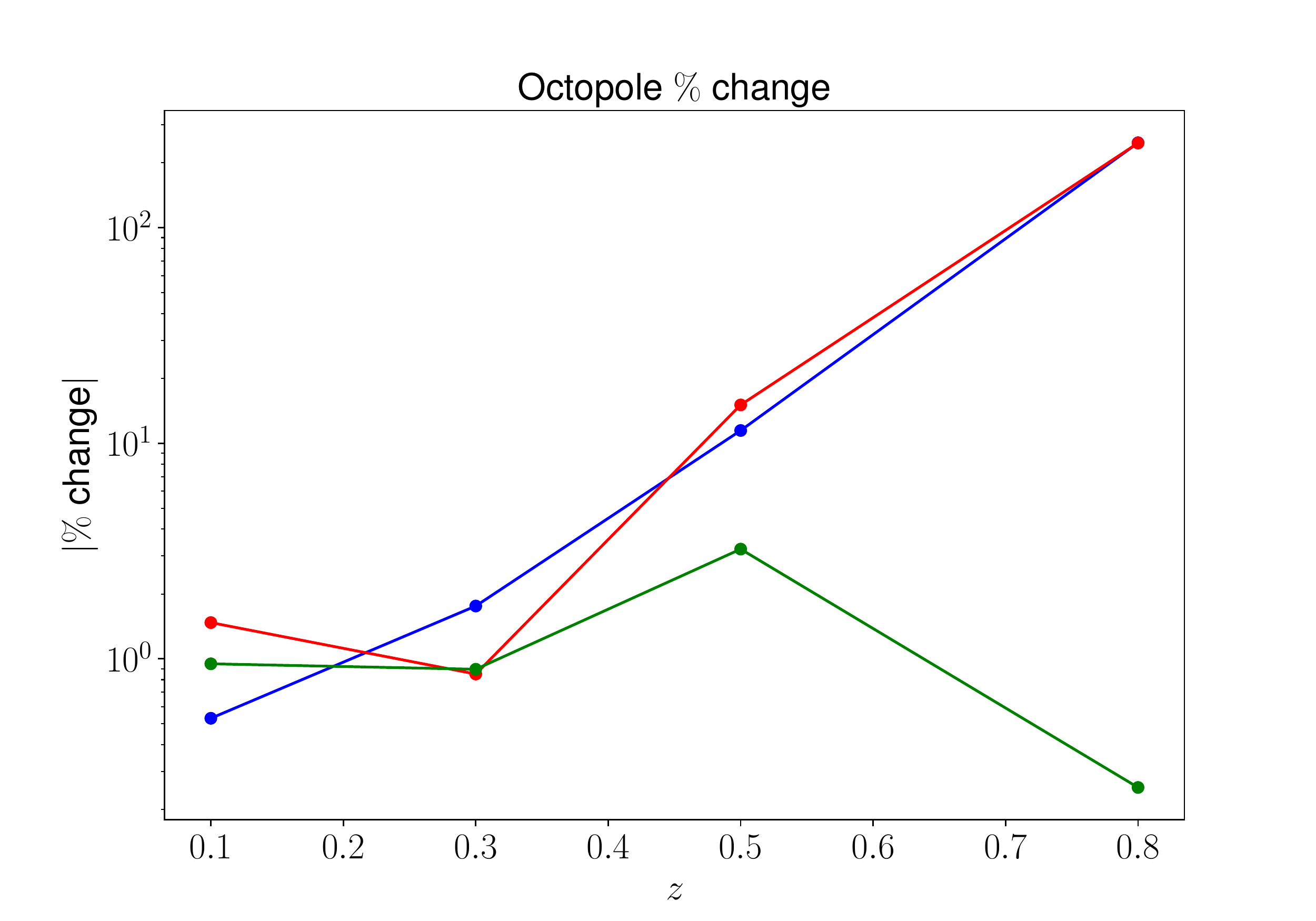}
   \caption{}
  \label{pc_oct}
\end{subfigure}
\caption{Absolute values of percentage change in the dipole and octopole under local screening and/or background modification as a function of redshift at a scale of $30$ Mpc. The percentage difference is defined relative to $\Lambda$CDM with the fiducial parameters of Table \ref{fid_cosmo}. The points, connected by straight lines, indicate the specific redshifts at which we perform the calculation. The legend is the same in the two panels.}
\label{pchange}
\end{figure*}

To summarise the effects of modified gravity, we show in Fig.~\ref{pchange} the variation of the percentage change with redshift in the range $0.1<z<0.8$, at a fiducial scale of $30$ Mpc. It is worth remembering that the octopole is always negative for our parameter choices, and the effect of screening is to reduce the dipole (hence the red line lying below the green in Fig.~\ref{pc_dip}). We see that the octopole is very sensitive to screening parameters at high redshift even with $s_B = 0.1$ (which determines the size of the octopole in our $\Lambda$CDM model). The change in dipole due to local screening is at most $\mathcal{O}(10\%)$ at high $z$, while the effect of background modification in non-screened modified gravity is roughly independent of redshift.

Current measurements of the dipole, for example from the LOWz and CMASS samples of the BOSS survey (e.g. \cite{Gaztanaga:2015jrs} fig. 7), have a signal to noise ratio (SN) of $<1$ in the dipole. In addition to the relativistic effects considered here, there are other terms that contribute to this and are in fact larger, including the wide angle and large angle effects. It is shown in \cite{Gaztanaga:2015jrs} that the current data from BOSS is only able to detect (at $\sim$2$\sigma$) the large angle effect, which is is a geometrical combination of the monopole and quadruple and hence contains no additional physical information. Thus the detection of relativistic effects in the dipole will require data from future surveys such as DESI and SKA. As the modifications due to screening are $\mathcal{O}(10 \%)$ in the dipole at high redshift one would need SN$\gtrsim$10 to detect them, while the detection of local screening at low redshift would require an additional order of magnitude improvement (along with more precise modelling of the bias parameters). While it will be challenging for future surveys to detect the dipole at SN=100 (e.g. it is forecast in \cite{Gaztanaga:2015jrs} that DESI will reach SN$\approx$7), it is shown in \cite{Hall:2016bmm} that a combination of SKA intensity maps and galaxy number counts can reach this sensitivity. Therefore it may be feasible for future surveys to probe both modified gravity in general, and screening in particular, through the dipole. To quantify the exact sensitivity to screening parameters, folding in uncertainties in growth rate and bias as well as cosmological variables such as $m_\nu, \Omega_m$ and $\sigma_8$, one could perform a Fisher forecast for next-generation experiments with cross-correlations of tracers. \emph{Any} measurement of the octopole may provide information on modified gravity.

Finally, it is worth noting that while we have computed the effect of the fifth force on the CF across a wide range of scales, in general we expect the force to be suppressed beyond the Compton wavelength of the field responsible for it. Given a Compton wavelength one can simply read off the change in the CF from Fig.~\ref{dip_oct} up to that scale, and assume a rapid transition back to $\Lambda$CDM beyond that. This applies only to the local fifth-force modification, however, and not the background. A completely self-consistent model would require solving the equation of motion for the field numerically given the mass distribution of the volume under consideration, while ensuring that the same fundamental theory parameters source both local screening and cosmic structure formation.

\section{Summary and future work}
\label{sec:conc}

We have calculated the effect of a screened fifth force on the parity-breaking correlation function (CF) obtained by cross-correlating two populations of galaxies that differ in properties relevant to their clustering. We show that this generates new terms in the dipole and octopole that are not present in $\Lambda$CDM, and in some cases neither in non-screened modified gravity theories. In particular, provided the magnification bias is the same between the two galaxy populations \emph{the octopole is only present under screening}. Should the octopole be detected it could provide a relatively clean probe of a screened fifth force.

The CF is also affected by cosmological modified gravity in the background, which alters the transfer function and growth rate. To model this we use a version of CAMB that has been modified \cite{Winther} to implement Hu-Sawicki $f(R)$, a canonical chameleon-screened theory. We find that Hu-Sawicki models with a fifth force on the scales in which we are interested ($10^{-6} \lesssim f_{R0} \lesssim 10^{-5}$) lead to deviations of $\mathcal{O}(10\%)$ in the dipole. To model the effect of screening we assume bright galaxies are completely screened in the Euler equation (i.e. feel GR), whereas faint galaxies feel the full fifth force. This is, of course, an approximation that is unlikely to be true in a cosmological setting, however it allows us to estimate the strength of the signal to screening.
For fifth-force strengths $\sim$10$-100\%$ of Newtonian gravity this leads to further changes in the dipole and octopole of a few percent at redshifts below $z = 0.3$, while for higher redshifts, e.g. $z = 0.8$, the dipole and octopole can change by $\sim 10-20\%$. We also show that uncertainties in the magnification and galaxy bias affect the dipole at the $\sim10\%$ level across the redshift range we consider, and therefore need to be known or modelled to this precision in order to extract information about screening.

Current state-of-the-art data from BOSS has signal to noise $<1$ in the dipole, and is not therefore able to detect these effects \cite{Gaztanaga:2015jrs}. However, upcoming DESI data will increase the signal to noise to $\sim$7, which will provide sensitivity to interesting modified gravity modifications to the cosmological background and local screening parameters at $z \approx 0.8$. The effect of screening at lower redshift may also be detectable using cross-correlations of multiple tracers, for example SKA intensity maps with galaxy number counts from Euclid or DESI \cite{Hall:2016bmm}. Another interesting prospect is the cross-correlation of galaxies with voids, which have negative bias and hence maximise the bias difference with the bright galaxy sample. This would however render the effect of screening further subdominant to the $\Lambda$CDM dipole. As the magnification and galaxy bias affect even multipoles of the CF as well, the best way to constrain the combination of bias parameters \emph{and} screening (which affects the odd multipoles) will be to do a joint inference on all multipoles simultaneously.

We have quantified modified gravity at the background level for the chameleon-screened Hu-Sawicki model of $f(R)$ only. Our analytic result in Eq.~\ref{eq:final}, however, holds for \emph{all} screened theories, including those that employ qualitatively different mechanisms such as Vainshtein. Even within the chameleon paradigm Hu-Sawicki $f(R)$ covers only a small fraction of the parameter space. In Appendix \ref{horndeski} we cast the general action for chameleon screening into Horndeski form. Therefore, a natural follow up would be to investigate the parity-breaking CF across the full chameleon (or more general) parameter space in the background, which could be achieved by implementing the general action in a modified gravity Boltzmann code such as HiClass \cite{hiclass}.

{It is worth recalling here the assumptions that go into our analytic calculation of the correlation function, which may limit its scope:}

\begin{itemize}

\item {Our main assumption is that the effect of screening can be accounted for by a simple modification to the acceleration of the form Eq.~\ref{FF}, with constant $\delta G$. This amounts to the approximation that the two sets of galaxies we are correlating are either completely screened or completely unscreened. In reality galaxies may be partly screened, and the fifth force may be sourced by only a fraction of the matter that sources the Newtonian force. This generically reduces $\delta G$ below $1/3$ in $f(R)$, and hence reduces the magnitude of $\xi^{(\mathcal{F})}$. To account for this fully one would need to solve the scalar field equations numerically given the density field surrounding the galaxies (e.g. \cite{Shao}).}

\item {We have shown that the halo/galaxy bias is degenerate with the strength of screening in the dipole. It is therefore important for the bias to be known or modelled accurately in order to extract the screening signal. An order $\mathcal{O}(10\%)$ change in the bias at high redshifts\footnote{This also depends on mass.} (which is expected in $f(R)$ theory, e.g.~\cite{Schmidt:2008tn}) will lead to a change in the dipole of a similar magnitude. Thus we must be able to model the galaxy bias to percent level precision in modified gravity in order to isolate the screening signal. We have attempted to account for this by taking the bias, as a function of mass, from $f(R)$ simulations \cite{Arnold:2018nmv}, and accounting for the change in bias self-consistently as a function of redshift by using the using the $f(R)$ growth factor. The octopole however is independent of halo/galaxy bias, making it a particularly robust probe of screened fifth forces.}

\item {Assembly bias will generically lead to differences in the bias of the two galaxy populations at fixed halo mass, and more generally to deviations from the Sheth--Tormen prediction. It is also a function of modified gravity, as halos tend to form earlier in cosmologies with larger $f_{R0}$. A more precise understanding of this phenomenon will aid in distinguishing bias differences from effects related to screening.}

\end{itemize}

{The effect of the environment in which the galaxies form (represented by the trees in Fig.~\ref{windy}), could also contribute to the breaking of parity. For example, as bright galaxies are likely to have more dust around them than faint galaxies, a faint galaxy in front of a bright one will appear brighter than an identical one behind \cite{Fang:2011hc, Bonvin:2013ogt}.}

{The fully self-consistent way to account for all of these effects is to run numerical simulation of structure formation under screened modified gravity (e.g. \cite{Winther:2015wla, Arnold:2013nfr, Puchwein:2013lza, Arnold_1, Arnold_2}) and then compute the parity-breaking correlation function directly from the resultant galaxy density field. The advantage of our analytic approach is that it brings out the physical processes underlying such parity breaking and hence reveals novel features such as the presence of an octopole, which is not typically calculated in simulations. More detailed numerical modelling than we have performed will in any case likely be necessary to extract, validate and interpret a signal from data.}

In summary, the effects of modified gravity in the parity-breaking CF could be probed in the near future with surveys such as DESI and SKA, with the best hope for constraining screening parameters coming from cross-correlation of tracers at multiple wavelengths. These analyses, {augmented by numerical simulations}, should be included in the fundamental physics agenda of large scale structure surveys in the coming decade.

\section*{Acknowledgments}
\noindent
We thank Shadab Alam, David Alonso, Emilio Bellini, Pedro Ferreira and Eva Mueller for useful discussions, {and two anonymous referees for comments which improved the manuscript.}
DK acknowledges financial support from ERC Grant No: 693024. 
HD is supported by St John's College, Oxford.

\bibliography{all_active}

\onecolumngrid

\newpage

\appendix

\section{Derivation of screened CF}\label{corr_funcs}

In this section we calculate the relativistic component of the two-point CF, $\xi^{rel(\mathcal{F})}$, in a theory with a screened fifth force:
\begin{equation}
	\xi^{rel(\mathcal{F})}(z, z', \theta) = \langle \Delta^{st}_B(z, \hat{\textbf{n}}) \Delta^{rel(\mathcal{F})}_F (z', \hat{\textbf{n}}') \rangle + \langle \Delta^{st}_F(z', \hat{\textbf{n}}') \Delta^{rel(\mathcal{F})}_B (z, \hat{\textbf{n}}) \rangle \label{corr_in}.
\end{equation}
We can substitute the expressions in Eqs (\ref{dst}, \ref{drel}) into Eq (\ref{corr_in}). We work in Fourier space and use the following convention for the Fourier transform of some function $f$:
\begin{equation}
	FT[f(x, \eta)] \equiv \frac{1}{(2\pi)^3} \int d^3k \ e^{-i \textbf{k} \cdot \textbf{x}} \ \mathbb{F} (\textbf{k}, \eta)
\end{equation}
We describe the Fourier transform of the density and velocity as
\begin{eqnarray}
	& & FT[\delta(x,\eta)] = \mathbb{D}(k, \eta) \nonumber \\
	& & FT[v(x,\eta)] = \mathbb{V}(k, \eta). \nonumber \\
\end{eqnarray}
These can be directly related to the transfer functions for the metric potential and the underlying initial metric perturbations $\Psi_i$. 
\begin{eqnarray}
	& & \mathbb{D}(\textbf{k}, \eta) = T_D(k, \eta) \Psi_i(\textbf{k}) \nonumber \\
	& & \mathbb{V}(\textbf{k}, \eta) = T_V(k, \eta) \Psi_i(\textbf{k}) \nonumber \\
	& & T_\Psi = T_\Phi = \frac{D(a)}{a} T(k) \nonumber \\
	& & T_D = - \frac{2a}{3 \Omega_m} \left( \frac{k}{\mathcal{H}_0} \right)^2 T_\Psi = - \frac{2}{3 \Omega} \left( \frac{k}{\mathcal{H}_0} \right)^2 D(a) T(k) \nonumber \\
	& & T_V = - \frac{\dot{T}_D}{k} = \frac{2a \mathcal{H}}{3 \Omega_m \mathcal{H}_0} \frac{k}{\mathcal{H}_0} \left[ T_\Psi + \mathcal{H}^{-1} \dot{T}_\Psi \right]  = \frac{2}{3 \Omega_m} \frac{\mathcal{H}}{\mathcal{H}_0} \frac{k}{\mathcal{H}_0} f(a)D(a)T(k)
\end{eqnarray}
where in the last line we have followed the standard convention, following \cite{Bonvin:2018ckp},  of splitting the time-dependent component of the transfer function into the linear growth factor $D(a)$ and the scale-dependent component into a time-dependent transfer function $T(k)$. 
In general modified gravity theories, this type of separation may not be possible as the growth can be scale dependent however we don't include that in this analysis and leave that to future works. 
We need only calculate one of the terms in Eq. (\ref{corr_in}) as the other will be related to this under $B \leftrightarrow F$, $z \leftrightarrow z'$, $\hat{\textbf{n}} \leftrightarrow \hat{\textbf{n}}'$. 
\begin{equation}
	\langle \Delta^{st}_B(z, \hat{\textbf{n}}) \Delta_F^{rel(\mathcal{F})} (z', \hat{\textbf{n}}') \rangle = \langle \Delta^{st}_B(z, \hat{\textbf{n}}) \Delta_F^{rel} (z', \hat{\textbf{n}}') + \langle \Delta^{st}_B(z, \hat{\textbf{n}}) \Delta_F^{\mathcal{F}} (z', \hat{\textbf{n}}') \rangle
\end{equation}
We compute each of these terms individually. 
\begin{eqnarray}
	& & \langle \Delta^{st}_B(z, \hat{\textbf{n}}) \Delta_F^{rel} (z', \hat{\textbf{n}}') \rangle =\mathcal{T}^{(1)} - \mathcal{T}^{(2)}  \\ 
	& & \mathcal{T}^{(1)} = \int \frac{d \ln k}{(2 \pi)^3} (k \eta_0)^{n_s -1} G(r') T_V(k,r') b_B T_D(k,r) \mathcal{I}^{(1)}  \\
	& & \mathcal{I}^{(1)} = \int d \Omega_k e^{i\textbf{k}( \textbf{x} - \textbf{x}')} (i \hat{\textbf{k}} \cdot \hat{\textbf{n}}')  \\
	& & \mathcal{T}^{(2)} = \int \frac{d \ln k }{(2 \pi)^3} (k \eta_0)^{n_s -1} G(r') T_V(k,r') \frac{k}{\mathcal{H}(r)} T_V(k,r) \mathcal{I}^{(2)}  \\
	& & \mathcal{I}^{(2)}\int d \Omega_k e^{i \textbf{k}( \textbf{x} - \textbf{x}')} (i \hat{\textbf{k}} \cdot \hat{\textbf{n}}') (\hat{\textbf{k}} \cdot \hat{\textbf{n}})^2
\end{eqnarray}
where we have defined 
\begin{equation}
	F(r) \equiv \frac{ \dot{\mathcal{H}}}{\mathcal{H}^2} + \frac{2}{r\mathcal{H}(r)} + 5s_B(r) \left( 1 - \frac{1}{r \mathcal{H}(r)} \right).
\end{equation}
To compute the angular integrals we use the following identities
\begin{eqnarray}
	e^{i \textbf{k} (\textbf{x}' - \textbf{x})} & = & e^{i d\textbf{k} \cdot \hat{\textbf{n}} } = 4 \pi \sum_{LM} j_L(kd) Y^*_{LM}(\hat{\textbf{k}}) Y_{LM}(\hat{\textbf{n}}) \nonumber \\
	\hat{\textbf{k}} \cdot \hat{\textbf{n}} & = & \frac{4 \pi}{3} \sum^1_{m = -1} Y^*_{1m} (\hat{\textbf{k}}) Y_{1m} (\hat{\textbf{n}}) \nonumber \\
	(\hat{\textbf{k}} \cdot \hat{\textbf{n}})^2 & = & \frac{8 \pi}{15} \sum_{m = -2}^2  Y^*_{2m}(\hat{\textbf{n}}) Y_{2m}(\hat{\textbf{k}}) + \frac{1}{3},
\end{eqnarray}
and we also make use of the \emph{Gaunt} integral formula
\begin{eqnarray}
	G^{l_1, l_2, l_3}_{m_1, m_2, m_3} & \equiv & \int d \Omega Y_{l_1, m_1}(\hat{n}) Y_{l_2, m_2}(\hat{n}) Y_{l_3, m_3} (\hat{n}) \nonumber \\
	& = & \sqrt{\frac{(2l_1+1)(2l_2+1)(2l_3+1)}{4 \pi}} \begin{pmatrix} l_1 & l_2 & l_3 \\ 0 & 0 & 0 \end{pmatrix} \begin{pmatrix} l_1 & l_2 & l_3 \\ m_1 & m_2 & m_3 \end{pmatrix} 
\end{eqnarray}
where we have defined the usual Wigner 3j symbol.
Now we compute the angular integrals as
\begin{eqnarray}
	\mathcal{I}^{(1)} & = & - 4 \pi \cos \alpha j_1(kd) \nonumber \\
	\mathcal{I}^{(2)} & = & 4 \pi \left[ - \frac{2}{5} \sin \alpha \sin \beta \cos \beta \left[ j_1(kd) + j_3(kd) \right] + \frac{3}{5} \cos \alpha \cos(2 \beta)  \left[ \frac{j_3}{2} - \frac{j_1}{3} \right] + \frac{1}{10} \cos \alpha \left[ j_3(kd) - j_1(kd) \right] \right].
\end{eqnarray}
Using these we can now compute the $k$ integrals
\begin{eqnarray}
	\mathcal{T}^{(1)} & = & \frac{2A_s}{9 \Omega^2 \pi^2} G(r') D(r)D(r') b_B(r) \cos \alpha \nu_1(d) f(r') \nonumber \\
	\mathcal{T}^{(2)} & = & \frac{2A_s}{9 \pi^2 \Omega^2_m} G(r') D(r) D(r') f(r) f(r') \left[ \frac{2}{5} \sin \alpha \sin \beta \cos \beta \left[ \nu_1(d) + \nu_3(d) \right] + \frac{3}{5} \cos \alpha \cos(2 \beta) \left[ \frac{\nu_3(d)}{2} - \frac{\nu_1(d)}{3} \right]  \right.\nonumber \\
	 & + & \left. \frac{1}{10} \cos \alpha \left[\nu_3(d) - 4\nu_1(d) \right] \right]. \nonumber \\
\end{eqnarray}
Thus the final answer for the two-point CF is 
\begin{eqnarray}
	\langle \Delta^{st}_B(z, \hat{\textbf{n}}) \Delta_F^{rel}(z', \hat{\textbf{n}}') \rangle & = & \frac{2 A_s G(r') D(r) D(r') f(r')}{9 \pi^2 \Omega_m^2} \left[ \frac{2}{5} \sin \alpha \sin \beta \cos \beta [\nu_1(d) + \nu_3(d)] + \frac{3}{5} \cos \alpha \cos(2 \beta) \left[ \frac{\nu_3(d)}{2} - \frac{\nu_1(d)}{3} \right] \right. \nonumber \\
	& & \left. + \frac{1}{10} \cos \alpha \left[ \nu_3(d) - 4 \nu_1(d) \right]  + b_B \nu_1(d) \cos \alpha\right]. \label{rel_temp1}
\end{eqnarray}
$\langle \Delta^{st}_F(z', \hat{\textbf{n}}') \Delta_B^{rel}(z, \hat{\textbf{n}}) \rangle$ is simply given by the relabelling of the indices and angles (which gives an rise to a sign difference in this term),
\begin{eqnarray}
	\langle \Delta^{st}_F(z', \hat{\textbf{n}'}) \Delta_B^{rel}(z, \hat{\textbf{n}}) \rangle & = & - \frac{2 A_s G(r) D(r') D(r) f(r)}{9 \pi^2 \Omega_m^2} \left[ \frac{2}{5} \sin \beta \sin \alpha \cos \alpha [\nu_1(d) + \nu_3(d)] + \frac{3}{5} \cos \beta \cos(2 \alpha) \left[ \frac{\nu_3(d)}{2} - \frac{\nu_1(d)}{3} \right] \right. \nonumber \\
	& & \left. + \frac{1}{10} \cos \beta \left[ \nu_3(d) - 4 \nu_1(d) \right]  + b_F \nu_1(d) \cos \beta\right]. \label{rel_temp2}
\end{eqnarray}
Next we compute the CF between the standard term for $B$ galaxies and the fifth-force term for $F$ galaxies. 
\begin{eqnarray}
	& & \langle \Delta^{st}_B(z, \hat{\textbf{n}}) \Delta_F^{\cal F}(z', \hat{\textbf{n}}') \rangle = A \int \frac{d^3k}{(2 \pi)3} e^{i \textbf{k} \cdot ( \textbf{x}' - \textbf{x})} \frac{(k \eta_0)^{n_s -1}}{k^3} \left[ b_B T_D(k,r) - \frac{k}{\mathcal{H}(r) }(\hat{\textbf{k}} \cdot \hat{\textbf{n}})^2 T_V(k,r) \right]  \times \nonumber \\
	& &  \left[ \zeta_F(r') i(\hat{\textbf{k}} \cdot{n}') \left( \mathcal{H}^{-1} (r') \dot{T}_v (k,r) + T_v(k,r') \right) \right] \nonumber \\
	& & \equiv  {\cal T}^{(3)} - \mathcal{T}^{(4)},
\end{eqnarray}
where $\zeta_F \equiv \frac{\delta G_F}{1 + \delta G_F}$ is the fifth-force sensitivity for faint galaxies. Here we define
\begin{eqnarray}
	\mathcal{T}^{(3)} & = & A \int \frac{d \ln k}{(2 \pi)^3} \zeta_F(r') \left[ \mathcal{H}^{-1}(r') \dot{T}_V(k,r') + T_V(k,r') \right] b_B T_D(k,r) \mathcal{I}^{(1)} \nonumber \\
	\mathcal{I}^{(1)} & \equiv  & \int d \Omega_k e^{i \textbf{k} (\textbf{x}' - \textbf{x})} (i \hat{\textbf{k}} \cdot \hat{\textbf{n}}') \nonumber \\
	\mathcal{T}^{(4)} & = & A \int \frac{ d \ln k}{( 2\pi)^3} \zeta_F(r') \left[ \mathcal{H}^{-1} (r') \dot{T}_V(k,r') + T_V(k,r') \right] \mathcal{H}^{-1}(r) k T_V(k,r) \mathcal{I}^{(2)} \nonumber \\
	\mathcal{I}^{(2)} & \equiv & \int d \Omega_k e^{- i \textbf{k} (\textbf{x}' - \textbf{x})} ( i \hat{\textbf{k}} \cdot \hat{\textbf{n}}') ( \hat{\textbf{k}} \cdot \hat{\textbf{n}})^2.
\end{eqnarray}
The other ingredients we need are the transfer functions
\begin{eqnarray}
	& & \dot{T}_V = \frac{2}{3 \Omega_m \mathcal{H}_0} \left( \frac{k}{\mathcal{H}_0} \right) \left[ \ddot{a} T_\Psi + \dot{T}_\Psi \left[ \ddot{a} \mathcal{H}^{-1} + \dot{a} - \dot{a} \dot{\mathcal{H}} \mathcal{H}^{-2} \right] + \dot{a} \mathcal{H}^{-1} \ddot{T}_\Psi \right] \nonumber \\
	& = & \frac{2}{3 \Omega_m} \left( \frac{k}{\mathcal{H}_0}\right) T(k) \left[ \frac{\dot{\mathcal{H}}}{\mathcal{H}_0} f(a)D(a) + \frac{\mathcal{H}(a)}{\mathcal{H}_0} \left( \dot{f}(a)D(a) + f(a)^2 D(a) \mathcal{H}\right)  \right],
\end{eqnarray}
where $ f(a) \equiv \frac{ d \ln D(a)}{d \ln a}$.  Further we can use $\dot{D} = fD\mathcal{H}$. We then put these into $\mathcal{T}^{(3)}$ and $\mathcal{T}^{(4)}$:
\begin{eqnarray}
	\mathcal{T}^{(3)} & = & \frac{2 A b_B \zeta_F \cos \alpha}{9 \pi^2 \Omega_m^2} D(r) D(r') \int \frac{d \ln k}{(2 \pi)^3} \frac{(k \eta_0)^{n_s-1}}{k^3} \left( \frac{k}{\mathcal{H}_0} \right)^3 T(k)^2 \left[ \frac{\mathcal{H}(r')}{\mathcal{H}_0} f(r') + \frac{\dot{\mathcal{H}(r')}}{\mathcal{H}_0} f + \frac{\mathcal{H}(r')}{\mathcal{H}_0} \dot{f}(r') + \frac{\mathcal{H}^2(r')}{\mathcal{H}_0(r')} f^2 \right] j_1(kd) \nonumber \\
	& = & \frac{2 \cos \alpha A b_B \zeta_F}{9 \pi^2 \Omega_m^2} M(r') \nu_1(d),
\end{eqnarray}
where we have defined 
\begin{eqnarray}
	& & \nu_\ell(d) \equiv \int \frac{dk}{k} (k \eta)^{n_s -1} \left( \frac{k}{\mathcal{H}_0} \right)^3 j_\ell(kd) T^2(k) \nonumber \\
	& & M(r') \equiv \frac{D^2(r')}{\mathcal{H}_0 \mathcal{H}} \left( \dot{\mathcal{H}(r')} f(r') + \mathcal{H}(r') \dot{f}(r') + f(r')^2 \mathcal{H}(r')^2 + f(r') \mathcal{H}(r')^2 \right).
\end{eqnarray}
Now we compute $\mathcal{T}^{(4)}$ as 
\begin{eqnarray}
	\mathcal{T}^{(4)} & = & A \int \frac{ d \ln k}{(2\pi)^3} \zeta_F(r') \left[ \frac{2}{3 \Omega_m} \left( \frac{k}{\mathcal{H}_0} T(k) \left[ \frac{\dot{\mathcal{H}}(r')}{\mathcal{H}_0} f(r') D(r') + \frac{\mathcal{H}(r')}{\mathcal{H}_0} \left( \dot{f}(r') D(r') + f(r')^2 D(r') \mathcal{H}(r') \right) \right] \right) \right] \nonumber \\
	& \times & b_B\left( \frac{k}{\mathcal{H}_0} \right) \left( \frac{k}{\mathcal{H}(r)} \right) \left[ \frac{2}{3 \Omega_m} \frac{\mathcal{H}(r)}{\mathcal{H}_0} f(r) D(r) T(k) \right] \nonumber \\
	& \times & 4 \pi \left[ - \frac{2}{5} \sin \alpha \sin \beta \cos \beta [j_1(kd)  + j_3(kd) ] + \frac{3}{5} \cos \alpha \cos(2 \beta) \left[ \frac{j_3(kd)}{2} - \frac{j_1(kd)}{3} \right] + \frac{1}{10} \cos \alpha \left[ j_3(kd) - j_1(kd) \right] \right] \nonumber \\ 
	& = & \frac{2 A \zeta_F f(r') M(r')}{9 \pi^2 \Omega_m^2 \mathcal{H}(r')} \left[ - \frac{2}{5} \sin \alpha \sin \beta \cos \beta(\nu_1(d) + \nu_3(d)) + \frac{3}{5} \cos \alpha \cos (2 \beta) \left( \frac{\nu_3(d)}{2} - \frac{\nu_1(d)}{3} \right) + \frac{1}{10} \cos \alpha (\nu_3(d) - \nu_1(d)) \right]. \nonumber \\
\end{eqnarray}
Putting these pieces together, we find
\begin{eqnarray}
	& & \langle \Delta^{st}_B(z,\hat{\textbf{n}}) \Delta^{(F)}_F(z', \hat{\textbf{n}}') \rangle = \frac{2A_sM(r') \zeta_F}{9 \pi^2 \Omega_m^2} \left[ b_B \cos \alpha \nu_1(d) + \frac{f(r')}{\mathcal{H}(r')} \left[ \frac{2}{5} \sin \alpha \sin \beta \cos \beta (\nu_1(d) + \nu_3(d))  \right. \right. \nonumber \\
	& & \left. \left. - \frac{3}{5} \cos \alpha \cos(2 \beta) \left( \frac{\nu_3(d)}{2} - \frac{\nu_1(d)}{3} \right) - \frac{1}{10} \cos \alpha (\nu_3(d) - \nu_1(d)) \right] \right]. \label{FF_temp1}
\end{eqnarray}
$\langle \Delta^{st}_F(z',\hat{\textbf{n}}') \Delta^{(F)}_B(z, \hat{\textbf{n}}) \rangle$ is given by the appropriate relabelling of the indices and angles:
\begin{eqnarray}
	& & \langle \Delta^{st}_F(z',\hat{\textbf{n}}') \Delta^{(F)}_B(z, \hat{\textbf{n}}) \rangle = -\frac{2A_sM(r) \zeta_B}{9 \pi^2 \Omega_m^2} \left[ b_F \cos \beta \nu_1(d) + \frac{f(r)}{\mathcal{H}(r)} \left[ \frac{2}{5} \sin \beta \sin \alpha \cos \alpha (\nu_1(d) + \nu_3(d))  \right. \right. \nonumber \\
	& & \left. \left. - \frac{3}{5} \cos \beta \cos(2 \alpha) \left( \frac{\nu_3(d)}{2} - \frac{\nu_1(d)}{3} \right) - \frac{1}{10} \cos \beta (\nu_3(d) - \nu_1(d)) \right] \right]. \label{FF_temp2}
\end{eqnarray}
By expanding (Eq.~\eqref{FF_temp1}, Eq.~ \eqref{FF_temp2}) and (Eq.~\eqref{rel_temp1}, Eq.~\eqref{rel_temp2}) to leading order in $d/r$ we obtain the expressions in Eq.~\eqref{eq:final} for $\xi^{rel}$ and $\xi^{(\mathcal{F})}$ respectively.

\section{General chameleon screening in Horndeski theory}\label{horndeski}

This section casts generic chameleon-screened scalar--tensor theories to Horndeski form. We anticipate that this will be useful for implementing a more general theory than $f(R)$ in the background.

The general action for a scalar--tensor theory that is immune to instabilities and has second order equations of motion is the Horndeski action, given by \cite{Horndeski, Bellini:2014fua}:
\begin{eqnarray}
	S & = & \int d^4x  \sqrt{-g} \left[ \sum_{i=2}^5 \mathcal{L}_i + \mathcal{L}_m[g_{\mu \nu}] \right] \nonumber \\
	\mathcal{L}_2 & = & G_2(\phi, X) \nonumber \\
	\mathcal{L}_3 & = & -G_3(\phi, X) \Box \phi \nonumber \\
	\mathcal{L}_4 & = & G_4(\phi, X) R + G_{4X} (\phi, X) \left[ (\Box \phi)^2 - \phi_{; \mu \nu} \phi^{; \mu \nu} \right] \nonumber \\
	\mathcal{L}_5 & = & G_5(\phi, X) G_{\mu \nu} \phi^{; \mu \nu} - \frac{1}{6} G_{5X}(\phi, X) \left[ (\Box \phi)^3  + 2 \phi_{; \mu}^\nu \phi_{;\nu}^\alpha \phi_{; \alpha}^\mu - 3 \phi_{; \mu \nu} \phi^{; \mu \nu} \Box \phi \right]. \label{horndeski_eq}
\end{eqnarray}
This is written in the Jordan frame, in which the Lagrangian components $\mathcal{L}_i$ determine the dynamics of the metric and the scalar field $\phi$. $X$ is the canonical kinetic term of a scalar field $- \frac{1}{2} g_{\mu \nu} \partial^\mu \phi \partial^\nu \phi$. 
The $G_i$ are free functions of the scalar and its kinetic term, and we denote their derivatives by $G_{iX} \equiv \partial_X G_i$.

Any scalar field minimally coupled to gravity has the action 
\begin{equation}
	\tilde{S} = \int d^4 x \sqrt{- \tilde{g}} \left[ \frac{M_{pl}^2}{2} \tilde{R} - \tilde{g}_{\mu \nu}\frac{1}{2} \partial^\mu \phi \partial^\nu \phi - V(\phi) - \mathcal{L}_m (g_{\mu \nu}) \right], 
\end{equation}
where variables are in the Jordan frame unless denoted by a tilde, in which case they are in the Einstein frame. In the Einstein frame, the scalar is decoupled from the metric and hence the gravitational part of the action is the same as in GR. Working in Planck units, we transform this action to the Jordan frame with the conformal transformation \cite{Wald}
\begin{eqnarray}
	\tilde{g}_{\mu \nu} &  = & e^{-2 \alpha \phi} g_{\mu \nu} \nonumber \\
	\tilde{g} & = & e^{-8 \alpha \phi} g \nonumber \\
	\tilde{R} & = & e^{2 \alpha \phi} \left[R - 6 \alpha^2 g_{\mu \nu} \partial^\mu \phi \partial^\nu \phi - 6 \alpha \Box \phi \right]
\end{eqnarray}
This yields the Jordan-frame action
\begin{equation}
	S = \int d^4x \sqrt{-g} \: \exp\left( -2 \alpha \phi \right) \: \left[ \frac{1}{2} R + 3 \alpha \Box \phi - \left( \frac{1}{2} + 3 \alpha^2\right) g_{\mu \nu} \partial^\mu \phi \partial^\nu \phi - \exp (-2 \alpha \phi) V(\phi) - \mathcal{L}_m (g_{\mu \nu}) \right].
\end{equation}
In this frame the scalar field has the Poisson equation $\Box \phi = \partial_\phi V(\phi) + \alpha \rho$.
The effective potential is then $V_\text{eff}(\phi) = V(\phi) + \exp(\alpha \phi) \rho$. To implement the chameleon mechanism $V(\phi)$ is chosen such that $V_\text{eff}$ has a sharp minimum, corresponding to high mass, in regions of high density, and a shallow minimum, corresponding to low mass, in regions of low density. A canonical example is $V(\phi) = \Lambda^{4+n}/\phi^n$ with $\Lambda$ an energy scale and $n$ an as-yet undetermined exponent \cite{Khoury:2013yya, Burrage}.\footnote{Note however that only some choices for $n$ result in chameleon screening. The mass is an increasing function of density, as required, if $n > 0$, $-1 < n < 0$ or $n$ is an even negative integer. $n = 0$ is simply a cosmological constant, $n = -1, -2$ does not make mass a function of density, and there is no minimum of $V_\text{eff}$ when $n = -3, -5, -7, ...$.}

Comparing to the Horndeski form (Eq (\ref{horndeski_eq})) we find: 
\begin{eqnarray}
	&G_2 = & - \exp\left( -4 \alpha \phi \right) V(\phi) + \exp\left( -2 \alpha \phi \right) X \left( 1 + 6 \alpha^2 \right) \nonumber \\
	&G_3 = & -3 \alpha \exp \left( -2 \alpha \phi \right) \nonumber \\
	&G_4 = & \frac{1}{2} \exp \left( -2 \alpha \phi \right) \nonumber \\
	&G_{4X} = & G_5 = G_{5X} = 0
\end{eqnarray}
We note that the recent neutron star merger that constrains the speed of gravitational waves to be the same of the speed of light implies $G_5, G_{4X} = 0$ \cite{Noller:2018eht} and thus our action is almost as general as possible given this constraint.
The $f(R)$ Hu-Sawicki model corresponds to the range $-1 < n < -1/2$ \cite{Burrage}, with $k=1$ corresponding to $n=-1/2$. Future work could explore the full parameter space of chameleon screening by implementing this action in a Boltzmann code such as HiClass \cite{hiclass}.

\end{document}